\documentclass[aps, prb, twocolumn, amsmath, amssymb, superscriptaddress, reprint, longbibliography]{revtex4-2}

\usepackage[utf8]{inputenc}
\usepackage[T1]{fontenc}
\usepackage[colorlinks=true,citecolor=blue]{hyperref}
\usepackage{graphicx}
\usepackage{units}
\usepackage[dvipsnames]{xcolor}

\usepackage{changes}
\definechangesauthor[name=Fred, color=CornflowerBlue]{F}
\definechangesauthor[name=HvH, color=Peach]{HvH}
\definechangesauthor[name=CB, color=WildStrawberry]{CB}
\definechangesauthor[name=HM, color=Emerald]{HM}
\definechangesauthor[name=PK, color=Orchid]{PK}
\definechangesauthor[name=BS, color=Bittersweet]{BS}
\definechangesauthor[name=MT, color=Mahogany]{MT}

\newcommand{\Tc}{T_\mathrm c}
\newcommand{\kB}{k_\mathrm B}
\newcommand{\angstrom}{\text{\AA}\vphantom{A}}
\newcommand{\TD}{\Theta_\mathrm D}

\usepackage{dcolumn} 

\newcommand{\Jp    }{J_\parallel}
\newcommand{\Js    }{J_\perp}
\newcommand{\Jd    }{J_\delta}

\newcommand{\Cv}[1]{J_{\parallel,#1}}
\newcommand{\Ch}[1]{J_{\perp,#1}}
\newcommand{\Cd}[1]{J_{\mathrm{d},#1}}
\newcommand{\CV    }{C_\mathrm{V}}
\newcommand{\CH    }{C_\mathrm{H}}
\newcommand{\CD    }{C_\mathrm{D}}
\newcommand{\CViii }{C_\mathrm{V3}}
\newcommand{\CViiii}{C_\mathrm{V4}}

\newcommand{\xip    }{\xi_\parallel}
\newcommand{\xis    }{\xi_\perp}
\newcommand{\xid    }{\xi_\delta}

\newcommand{\alp    }{a_\parallel}
\newcommand{\als    }{a_\perp}
\newcommand{\ald    }{a_\delta}

\begin{document}


\title{Critical behavior of the dimerized Si(001) surface: Continuous order-disorder phase transition in the two-dimensional Ising universality class}




\author{Christian Brand}
\email[Corresponding author: ]{christian.brand@uni-due.de}
\affiliation{Faculty of Physics, University of Duisburg-Essen, 47057 Duisburg, Germany}

\author{Alfred Hucht}
\affiliation{Faculty of Physics, University of Duisburg-Essen, 47057 Duisburg, Germany}
\affiliation{Center for Nanointegration (CENIDE), University of Duisburg-Essen, 47057 Duisburg, Germany}

\author{Hamid Mehdipour}
\affiliation{Faculty of Physics, University of Duisburg-Essen, 47057 Duisburg, Germany}

\author{Giriraj Jnawali}
\altaffiliation[Present address: ]{Department of Physics, University of Cincinnati, Cincinnati, OH 45221, USA}
\affiliation{Faculty of Physics, University of Duisburg-Essen, 47057 Duisburg, Germany}

\author{\firstname{Jonas D.} Fortmann}
\affiliation{Faculty of Physics, University of Duisburg-Essen, 47057 Duisburg, Germany}

\author{Mohammad Tajik}
\affiliation{Faculty of Physics, University of Duisburg-Essen, 47057 Duisburg, Germany}

\author{R\"{u}diger Hild}
\affiliation{Faculty of Physics, University of Duisburg-Essen, 47057 Duisburg, Germany}

\author{Bj\"{o}rn Sothmann}
\affiliation{Faculty of Physics, University of Duisburg-Essen, 47057 Duisburg, Germany}
\affiliation{Center for Nanointegration (CENIDE), University of Duisburg-Essen, 47057 Duisburg, Germany}

\author{Peter Kratzer}
\affiliation{Faculty of Physics, University of Duisburg-Essen, 47057 Duisburg, Germany}
\affiliation{Center for Nanointegration (CENIDE), University of Duisburg-Essen, 47057 Duisburg, Germany}

\author{Ralf Sch\"{u}tzhold}
\affiliation{Institute of Theoretical Physics, Dresden University of Technology, 01062 Dresden, Germany}
\affiliation{Helmholtz-Zentrum Dresden-Rossendorf, 01328 Dresden, Germany}

\author{Michael \surname{Horn-von Hoegen}}
\affiliation{Faculty of Physics, University of Duisburg-Essen, 47057 Duisburg, Germany}
\affiliation{Center for Nanointegration (CENIDE), University of Duisburg-Essen, 47057 Duisburg, Germany}

\date{\today}


\begin{abstract}
The critical behavior of the order-disorder phase transition in the buckled dimer structure of the Si(001) surface is investigated both theoretically by means of first-principles calculations and experimentally by spot profile analysis low-energy electron diffraction (SPA-LEED). 
We use density functional theory (DFT) with three different functionals commonly used for Si to determine the coupling constants of an effective lattice Hamiltonian describing the dimer interactions.
Experimentally, the phase transition from the low-temperature $c(4 {\times} 2)$- to the high-temperature $p(2 {\times} 1)$-reconstructed surface is followed through the intensity and width of the superstructure spots within the temperature range \unit[78-400]{K}.
Near the critical temperature $\Tc = \unit[190.6]{K}$, we observe universal critical behavior of spot intensities and correlation lengths, which falls into the universality class of the two-dimensional (2D) Ising model.
From the ratio of correlation lengths along and across the dimer rows we determine effective nearest-neighbor couplings of an anisotropic 2D Ising model, $\Jp = \unit[(-24.9 \pm 0.9_\mathrm{stat} \pm 1.3_\mathrm{sys})]{meV}$ and $\Js = \unit[(-0.8 \pm 0.1_\mathrm{stat})]{meV}$. 
We find that the experimentally determined coupling constants of the Ising model can be reconciled with those of the more complex lattice Hamiltonian from DFT when the critical behavior is of primary interest.
The anisotropy of the interactions derived from the experimental data via the 2D Ising model is best matched by DFT calculations using the PBEsol functional. 
The trends in the calculated anisotropy are consistent with the surface stress anisotropy predicted by the DFT functionals, pointing towards the role of surface stress reduction as a driving force for establishing the $c(4 {\times} 2)$-reconstructed ground state.
\end{abstract}

\maketitle


\section{Introduction}
\label{sec:Introduction}

\par
Materials with reduced dimensions often exhibit energetically close-lying metastable states due to combined instabilities of the electron and the lattice system.
Examples in one dimension (1D) include the formation of charge density waves, doubling of periodicity driven by a Peierls distortion, or symmetry breaking through a Jahn-Teller instability \cite{Ahn:PRL93.106401, Kumpf:PRL85.4916, wippermann2010entropy, schmidt2012si}. 
In purely 1D structures, these instabilities are predicted to exist only for $T = \unit[0]{K}$.
Even smallest thermal excitations drive the system into the high temperature state.
Only those interactions mediated by higher dimensions, e.g., the coupling of parallel 1D atomic wires perpendicular to the wire direction, stabilize the instability at finite temperatures.
Examples for such systems are adsorbate-induced atomic wires such as Si(553)-Au \cite{erwin2010intrinsic, Hafke:PRL124.016102, Mamiyev:PRL126.106101, snijders2010colloquium}, Si(557)-Pb \cite{Czubanowski:NJoP9.338, Pfnuer:SufSci643.79, Tegenkamp:PRL109.266401, Brand:NatComm6}, Ge(110)-Pt \cite{zhang2016structural} and Si(111)-In \cite{yeom1999instability, frigge2017optically}.

\par
A truly famous example of such a 1D structure with weak coupling in the second dimension is the bare Si(001) surface exhibiting a so-called dimer reconstruction \cite{dabrowski2000silicon, farnsworth1958application, ren2016origin}.
The bulk-terminated Si surface exhibits two half-filled dangling bonds per atom which makes the Si(001) surface electronically unstable.
Since thermodynamics always tends to minimize the surface free energy, these dangling bonds form dimers, leading to a doubling of the periodicity along the dimer axis and thus forming a reconstruction on the Si(001) surface \cite{schlier1959structure, Chadi:PRL43.43, Tromp:PRL55.1303}. 
These dimers, which are alternately buckled at low temperatures \cite{weakliem1990subpicosecond}, are arranged in parallel rows.
At low temperatures, the adjacent dimer rows are arranged in an anti-phase registry, leading to a $c(4 {\times} 2)$ reconstruction as the ordered ground state \cite{wolkow1992direct, zhao1986atomic}.
The buckling orientation can be formally assigned to an up- or down-spin, i.e., a two-dimensional (2D) antiferromagnetic order arises.
This system is known for the experimental observation of a continuous order-disorder phase transition that occurs at $\Tc \approx \unit[200]{K}$ from the $c(4 {\times} 2)$ to the $p(2 {\times} 1)$ high-temperature state \cite{Tabata:SurfSci179.L63}.
Similar to silicon, the Ge(001) surface also exhibits such a structural phase transition from a $c(4 {\times} 2)$ to a $p(2 {\times} 1)$ reconstruction at slightly higher temperature \cite{Zandvliet:PhysRep388.1, Zandvliet:SSC78.455, Zandvliet:JPCM:3.409, Lucas:PRB47.10375, Cvetko:SurfSci447.L147}.
Kevan \textit{et al.} \cite{Kevan:PRB32.2344} suggested that this order-disorder phase transition occurs via a two-step process for Ge(001).
Details of the phase transition on Ge(001) can be found elsewhere 
\cite{Zandvliet:PhysRep388.1}.

\par
Although the order-disorder phase transition for the dimerized Si(001) surface has been known for decades, the critical behavior has yet not been studied in detail.
Previous studies were constrained by the limited instrumental resolution of low-energy electron diffraction (LEED) and the pinning of the phase transition by the extreme sensitivity of the surface to adsorbates or the omnipresent atomic steps that prevent large-scale spatial fluctuations \cite{Kubota:PRB49.4810, Murata:PT53.125, Tabata:SurfSci179.L63, Matsumoto:PRL90.106103}.
Also, experimental findings such as the observation of streaklike diffuse intensity \cite{Kubota:PRB49.4810} could not be explained by theory \cite{Inoue:PRB49.14774}. 

\par
Very recently, we showed that the Si(001) order-disor\-der phase transition can be described by an anisotropic 2D Ising model with $\Tc = \unit[190.6]{K}$ \cite{Brand:PRL130.126203}.
The critical behavior of the order parameter, the fluctuations and the correlation lengths were analyzed in the framework of the 2D Ising universality class \cite{Kadanoff:Physics.2.263, Fisher:RevModPhys.70.653} and were mapped onto the exactly solved nearest-neighbor (NN) 2D Ising model \cite{Onsager:PR65.117, McCoy+Wu:2D.Ising}.
This mapping onto renormalized effective NN couplings was justified \textit{a posteriori} by the very large coupling anisotropy ratio \cite{Brand:PRL130.126203}.
The experimentally determined coupling energies $\Jp = \unit[-24.9]{meV}$ and $\Js = \unit[-0.8]{meV}$ along and across the dimer rows, respectively, were in agreement with some previously conducted \textit{ab initio} density functional theory (DFT) calculations, while other such calculations deviated from the experimental results.
To elucidate the microscopic interactions involved, we performed comprehensive DFT calculations with three different exchange-correlation functionals.
Our paper thus helps to resolve the long-standing debate as to why some theoretical predictions of $\Tc$ (e.g. $\Tc = \unit[316]{K}$ \cite{Inoue:PRB49.14774}) differ from the experimental value. 

\par
This paper is organized as follows: After summarizing the current understanding of the energetic hierarchy of the Si(001) surface structure in Sec.~\ref{sec:dimerized}, we review the techniques for setting up lattice Hamiltonians for the Si(001) surface and present our interaction parameters calculated from DFT in Sec.~\ref{sec:DFT}.
We then argue that, as far as the critical behavior close to the order-disorder transition is concerned, the lattice Hamiltonian can be reduced to an anisotropic 2D Ising Hamiltonian with two effective parameters (Sec.~\ref{sec:Ising}).
The analytically known scaling laws of the 2D Ising model allow us to precisely determine $\Tc$ and the ratio of the interaction parameters.
In Sec.~\ref{sec:experimental} the experimental data analysis used in Ref.~\cite{Brand:PRL130.126203} is explained in more detail.
Finally, we conclude and give an outlook on possible future work.
In the \hyperref[sec:Appendix]{Appendix~\ref*{sec:Appendix}}, we compare the results of our DFT calculations with previous studies.


\section{Dimerized Si(001) surface}
\label{sec:dimerized}

\begin{figure*}[ht]
\centering
\includegraphics[width=1.000\textwidth]{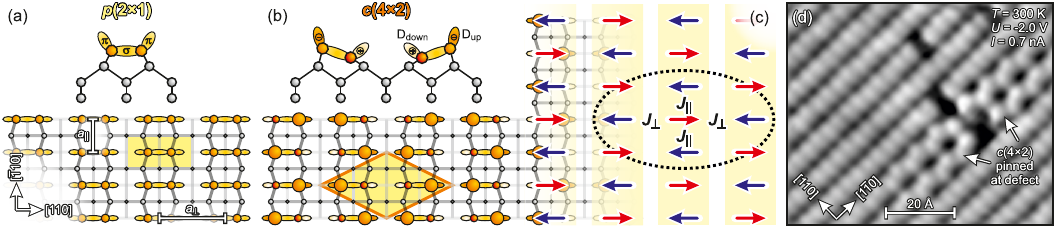}
\caption{
\textbf{Dimer reconstruction of Si(001).}
Atomic structure models of the Si(001) surface for (a) the symmetric $p(2 {\times} 1)$ and (b) the $c(4 {\times} 2)$ reconstruction.
(c) Spin model describing the arrangement of alternation of dimer buckling.
Effective interaction parameters $\Jp$ and $\Js$ along and across the dimer rows are indicated.
(d) STM image recorded at $T = \unit[300]{K}$, $U = \unit[-2.0]{V}$, $I = \unit[0.7]{nA}$ (occupied states) exhibiting buckled dimers in the vicinity of a surface defect. 
}
\label{Fig.Structure}
\end{figure*}

\par
The bare dimerized Si(001) surface exhibits a rich hierarchy of structural motifs that minimize the surface free energy \cite{Ramstad:PRB51.14504}.
The dimer formation is facilitated by the energy minimization of the bulk-terminated Si surface, where each surface atom has two half-filled dangling bonds.
This unstable structure results from the directional covalent bonds of the underlying diamond crystal structure.
Each dimer is composed of two Si surface atoms bonded to each other, resulting in a twofold periodicity along the dimer axis.
As will be seen in Sec.~\ref{sec:experimental}, the dimers form parallel dimer rows giving rise to a strong geometric and electronic anisotropy of the surface with a $p(2 {\times} 1)$ unit cell as sketched in Fig.~\ref{Fig.Structure}(a).
The corresponding LEED pattern reflects the doubling of the periodicity by additional spots halfway between the integer order spots, as shown in Fig.~\ref{Fig.Patterns}(a) below.
Each dimer reduces the number of half-filled dangling bonds by a $\sigma$-like bond resulting from the overlap of half-filled dangling bond orbitals as sketched in Fig.~\ref{Fig.Structure}(a).
This fully occupied $\sigma$-like bond reduces the lateral distance between the two Si surface atoms from \unit[3.84]{\angstrom} to \unit[2.23]{\angstrom} \cite{wandelt2012surface}.
The associated bending of the backbonds from the top Si surface atoms by $\approx 20^\circ$ brings the two remaining dangling bonds almost parallel to each other, allowing for further energy reduction by forming a $\pi$-like bond.
The energy gain of $\approx \unit[1.8]{eV}$ per $p(2 {\times} 1)$ unit cell due to the saturation of the dangling bonds overcompensates by far the energy cost of the unfavorable bending of the backbonds.
Due to this large gain of energy the Si(001) surface remains dimerized up to temperatures of at least $1200^\circ$C \cite{fukaya2003phase}.

\par
The symmetric dimer phase is found above \unit[900]{K} \cite{fukaya2003phase}, however, is instable to lateral and vertical buckling and can further reduce its energy by Jahn-Teller distortion: One of the dimer atoms moves inward toward the bulk, the other outward toward the vacuum as sketched in Fig.~\ref{Fig.Structure}(b) \cite{Chadi:PRL43.43}.
Such buckling does not require changes of bond lengths, which are almost conserved while the bond angles change drastically until the dimer is tilted by $18^\circ$ with respect to the surface plane \cite{Over:PRB55.4731, Felici:SurfSci375.55, Shirasawa:SurfSci600.815, Pillay:SurfSci554.150, Ramstad:PRB51.14504, Guo:JPhysChemC118.25614, Dabrowski:ASS56-58.15} as sketched in Fig.~\ref{Fig.Structure}(b).
This process is accompanied by a charge transfer of $\approx 0.1e_0$ ($e_0$ is the elementary charge) from the down-dangling bond $D_\mathrm{down}$ to the energetically more favorable $s$-like up-dangling bond $D_\mathrm{up}$ \cite{Landemark:PRL69.1588}.
This charge transfer from the lower to the upper dangling bond causes the ionic character of the surface atoms.
The repulsive force between equally charged dimer atoms induces an alternation of the dimer buckling direction along each dimer row.
Each negatively charged up-atom is now surrounded by four positively charged down-atoms reducing the electrostatic repulsion between the ionic dangling bonds.
Ramstad \textit{et al.} used first-principles calculations to determine the reduction of surface energy due to alternatively buckled dimers by another \unit[50]{meV/dimer} \cite{Ramstad:PRB51.14504}.

\par
There is still one degree of freedom left for the arrangement of the alternately buckled dimer rows.
The registry of buckling in neighboring dimer rows could be either in-phase as for the asymmetric $p(2 {\times} 2)$ reconstruction or phase shifted by one lattice spacing along the dimer rows resulting in the $c(4 {\times} 2)$ reconstruction as sketched in Fig.~\ref{Fig.Structure}(b) and observed in LEED as shown in Fig.~\ref{Fig.Patterns}(b) below.
Due to electrostatic interaction and minimization of surface stress, the $c(4 {\times} 2)$ structure is slightly (by \unit[3]{meV/dimer} according to the calculations of Ref.~\cite{Ramstad:PRB51.14504}) more favorable than the $p(2 {\times} 2)$ structure, where the buckling in adjacent dimer rows occurs in-phase.
This is supported by LEED above $\approx \unit[45]{K}$, where only faint intensity of the $(1/2~1/2)$ spot was reported \cite{Kubota:PRB49.4810, Murata:PT53.125, Yoshida:PRB70.235411}.
However, small portions of the surface may exhibit the $p(2 {\times} 2)$ reconstruction in coexistence with the $c(4 {\times} 2)$ reconstruction at very low temperatures below $\approx \unit[45]{K}$ \cite{Yoshida:PRB70.235411}, where the long-range order of the buckled dimers in the $c(4 {\times} 2)$ reconstruction can still be lifted locally by the creation and motion of antiphase translational domain boundaries (so-called phasons) \cite{Pennec:PRL96.026102, Shigekawa:JJAP35.L1081, Yokoyama:PRB61.R5078} and/or by electron doping by the probe \cite{Hata:PRL89.286104, Shirasawa:PRL94.195502, Mizuno:PRB69.241306, Seino:PRL93.036101, Kawai:JPSJ76.034602, Sagisaka:PRL91.146103, Sagisaka:PRB.71.245319}.

\par
D{\k{a}}browski \textit{et al.} determined an energy barrier of \unit[90]{meV} between the asymmetric left- and right-tilted states of the $p(2 {\times} 1)$ phase \cite{Dabrowski:ASS56-58.15}.
Their calculations exclude any metastable state as function of buckling angle, i.e., there exists no symmetric dimer configuration without buckling.
At room temperature, however, the dimers appear as a symmetric $p(2 {\times} 1)$ [Fig.~\ref{Fig.Structure}(a)] in scanning tunneling microscopy (STM) due to fast dynamical flipping motion between left- and right-tilted configurations at a frequency of $\unit[10^{11}]{s^{-1}}$ \cite{Dabrowski:ASS56-58.15}.
Figure~\ref{Fig.Structure}(d) shows a room temperature STM image at negative sample bias, i.e., imaging the occupied states of the Si(001) surface.
The $p(2 {\times} 1)$-reconstructed dimer rows are clearly visible.
Dark areas represent missing dimer defects.
In the vicinity of an extended defect the thermally activated flipping motion of the buckled dimers is quenched in two adjacent dimer rows, locally exhibiting the $c(4 {\times} 2)$ reconstruction.
The transition between the $c(4 {\times} 2)$ ground state and the $p(2 {\times} 1)$ reconstruction observed at room temperature is also described as an order-disorder transition.


\section{First-principles calculations}
\label{sec:DFT}


\subsection{DFT methods}
\label{sec:DFT-Method}

\begin{figure*}[ht]
\centering
\includegraphics[scale=1.000]{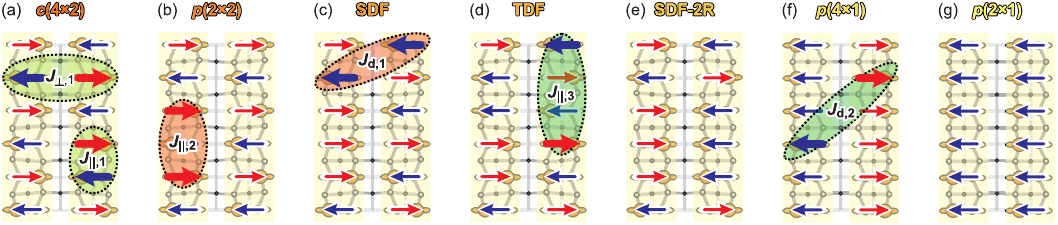}
\caption{\textbf{Schematics of dimer tilting patterns of the Si(001) surface.}
Seven different dimer tilting patterns of the Si(001) surface were studied in DFT modeling to determine up to six interaction parameters between the Si dimers (denoted by the highlighted arrows in the ellipsoids).
Here, SDF, TDF, and SDF-2R stand for single dimer flipped, twin dimers flipped (in one row), and single dimer flipped in adjacent rows, respectively.
Heads and tails of arrows denote the up- and down-dangling bonds of the tilted surface Si dimers as introduced is Fig.~\ref{Fig.Structure}(c).
}
\label{Fig.DFTstructure}
\end{figure*}

\par
Our goal is to describe the thermodynamics of the Si(001) surface by a lattice Hamiltonian.
Since the potential energy surface of the Si dimers has two clearly distinct minima and a saddle point in between \cite{dabrowski2000silicon, Dabrowski:ASS56-58.15, Healy:PRL.2001}, it is appropriate to model the buckling angle by Ising spins $\sigma_{i,j} = \pm 1$ corresponding to the two buckling orientations 'right' or 'left' of the Si dimers as sketched in Fig.~\ref{Fig.Structure}(c). 
The energetics of the Si(001) surface, including low-energy excitations of the Si dimer pattern, are described by a generalized anisotropic 2D Ising model on a rectangular lattice.
We use DFT calculations as implemented in the Quantum Espresso (QE) simulation package \cite{Giannozzi_2009} to determine the coupling parameters in this model.
QE is a periodic DFT computational suite that implements plane waves for the expansion of electronic wave functions.

\par
Here, we compare three different functionals to treat the exchange-correlation interactions of electrons, the local density approximation (LDA) as parametrized by Perdew and Zunger (PZ) \cite{prb-perdew-81}, and the generalized gradient approximations (GGAs) in the parametrizations PBE \cite{prl-perdew-96} and PBEsol \cite{prl-perdew-2008}.
The ion-valence electron interactions are treated using norm-conserving pseudo-potentials of Hartwigsen-Goedecker-Hutter type \cite{prb-hartwigsen-98} when PZ and PBE exchange-correlation functionals are used, and an ultra-soft pseudo-potential \cite{prb-vanderbilt-90} when the PBEsol functional is used.
For geometry relaxation the convergence thresholds of the total energy and each component of forces on each atom were set at \unit[0.01]{meV} and \unit[1]{meV/\angstrom}, respectively.
The calculations were performed at $T = \unit[0]{K}$.

\par
For these calculations a $(6 {\times} 4)$ supercell of the Si(001) surface with two Si dimer rows was constructed with eight Si atomic layers. 
The lateral dimensions of the supercell were set according to the bulk lattice constant of Si, which was determined individually for each functional. 
The 
supercell 
included a vacuum region of \unit[15]{\angstrom} to avoid artificial interactions between the slab and its image replicas normal to the surface.
In addition, total energy calculations for fixed geometries were carried out for a plane-wave cut-off of \unit[75]{Ry}, following a total energy convergence test of the cut-off energy.
To carry out total energy integration over the Brillouin zone of the supercell, a Monkhorst-Pack mesh \cite{MoPa76} of $3 {\times} 5 {\times} 1$ was used after fully examining the convergence of the total energy with respect to the size of the $k$ mesh.
The dangling bonds of the bottom-most Si atoms were saturated using H atoms and the Si and H atoms in these layers were kept fixed throughout the geometry relaxation.

\par
In the present paper, seven different tilting patterns of the Si(001) surface were modelled in order to obtain the coupling constants of the lattice Hamiltonian, which, as will be shown, describes the short-ranged dimer interactions and their contributions to the total energy of the Si(001) surface.
Figure~\ref{Fig.DFTstructure} displays the modelled tilting patterns.
Among the Si(001) surface tilting patterns, only the $p(2 {\times} 1)$ and $p(4 {\times} 1)$ reconstructions do not have an alternation of the buckling along a Si dimer row.
As will be shown, this translational symmetry along the Si dimer row introduces an energy penalty that makes these two tilting patterns energetically less favorable.
Also, according to previous experimental reports on the existence of domain boundaries (phasons) on the Si(001) surface \cite{Pennec:PRL96.026102}, we modelled $c(4 {\times} 2)$ patterns in which the domain boundaries exist in one row due to flipping of a single dimer in one row (SDF) or two dimers in two adjacent rows (SDF-2R), see Fig.~\ref{Fig.DFTstructure}.
We also included a tilting pattern with two domain boundaries formed by flipping of two dimers (TDF) in a single Si dimer row, see Fig.~\ref{Fig.DFTstructure}.


\subsection{Lattice Hamiltonian}
\label{sec:DFT-Hamiltonian}

In early studies \cite{Ihm:PRL51.1872, Alerhand:PRB35.5533, Inoue:PRB49.14774}, researchers included at least three short-range two-spin interactions $\Cv{1}$, $\Ch{1}$, and $\Cd{1}$, along, across, and diagonal to the Si dimer rows in their Hamiltonians, see Fig.~\ref{Fig.DFTstructure}.  
Later, conditional longer-range multispin interactions ($\CViii$ and $\CViiii$) were added \cite{Pillay:SurfSci554.150, Xiao:PRM3.044410}, which contribute only when three or four adjacent dimers are tilted in the same direction, see the \hyperref[sec:Appendix]{Appendix~\ref*{sec:Appendix}} for a discussion. 
In addition, a second-nearest-neighbor (NNN) interaction $\Cv{2}$ as well as a four-spin plaquette interaction were first introduced by Ihm {\it et al.} \cite{Ihm:PRL51.1872}, denoted as $U$ and $F$ in their Hamiltonian.

\par
In this paper, we demonstrate that it is sufficient to solely use two-spin interactions to account for the most important two-, three-, and four-dimer interactions within two dimer rows in our DFT model.
However, as the couplings along the dimer rows are much stronger than across the dimer rows \cite{Brand:PRL130.126203}, in addition to the nearest-neighbor (NN) interactions we allow for NNN and third-nearest-neighbor (NNNN) interactions in the direction along the dimer rows.
Thus, the generalized Ising model Hamiltonian used here reads
\begin{align}\label{eq:IsingHamiltonian_DFT}
\mathcal{H} = 
&-\Ch{1} \sum_{i,j} \sigma_{i,j} \sigma_{i+1,j} 
 -\sum_{r=1}^{3}\Cv{r} \sum_{i,j} \sigma_{i,j} \sigma_{i,j+r}
 \nonumber\\
&-\sum_{r=1}^{2}\Cd{r} \sum_{i,j} \sigma_{i,j} (\sigma_{i+1,j+r} + \sigma_{i+1,j-r})\,.
\end{align}
In the direction across the dimer rows longer-ranging interactions than $\Ch{1}$ are neglected since they turned out to be reasonably small.
For comparison with previous studies, we also explored two approaches that simply add $\CViii$, or both $\CViii$ and $\CViiii$ instead of $\Cv{2}$, $\Cv{3}$, and $\Cd{2}$.
These results are summarized in the \hyperref[sec:Appendix]{Appendix~\ref*{sec:Appendix}}.

\par
The large correlation length anisotropy present in the considered system and the resulting strong antiparallel short-range correlations in parallel direction \cite{Brand:PRL130.126203} gives rise to renormalized effective NN couplings of the form
\begin{subequations}\label{eqs:Jps}
\begin{align}
    \Jp &= \Cv{1} -   \Cv{2} +   \Cv{3}\,, \label{eq:Jp} \\
    \Js &= \Ch{1} - 2 \Cd{1} + 2 \Cd{2}\,, \label{eq:Js}
\end{align}
\end{subequations}
which will be used in the following.

\par
In the present paper, seven dimer tilting patterns of the Si(001) surface with varying thermal stability were studied, see Fig.~\ref{Fig.DFTstructure}.
The coupling parameters of the Hamiltonian given in Eq.~\eqref{eq:IsingHamiltonian_DFT} were determined by solving a system of linear equations corresponding to the tilting patterns considered.
Subtracting the energy of the reference ground state $c(4 {\times} 2)$, we find
\begingroup
\allowdisplaybreaks
\begin{subequations}
\begin{align}
E_{c(4{\times}2)} & = 0\,, \label{eq:E_c4x2}\\
E_{p(2{\times}2)} & = -24 \Js\,, \label{eq:E_p2x2}\\
E_\mathrm{SDF}    & = -4  ( \Jp + \Js )\,, \label{eq:E_SDF}\\
E_\mathrm{TDF}    & = -8  ( \Jp + \Js - \tfrac 1 2 \Cv{1})\,, \label{eq:E_TDF}\\
E_\mathrm{SDF-2R} & = -8  ( \Jp + \Js -            \Ch{1})\,, \label{eq:E_SDF-2R}\\
E_{p(4{\times}1)} 
 & = -24 ( \Jp - 2 \Cd{1} + \Cv{2})\,, \label{eq:E_p4x1}\\
E_{p(2{\times}1)} & = -24 ( \Jp + \Js + 2 \Cd{1} + \Cv{2})\,, \label{eq:E_p2x1} 
\end{align}%
\label{eqs:DFTTotalEnergies}%
\end{subequations}
\endgroup
where the quantities on the left-hand sides are the DFT total energies of the dimer tilting patterns in the $(6 {\times} 4)$ supercell.
As Eqs.~\eqref{eqs:DFTTotalEnergies} are written in terms of the renormalized effective NN couplings $\Jp$ and $\Js$ from Eq.~\eqref{eqs:Jps}, they can be solved successively as discussed below.


\subsection{DFT results on lattice Hamiltonian}
\label{sec:DFT-results}


\begin{table}[b]
\caption{
\textbf{DFT results on energetics of the tilting patterns.}
Relative energies of the tilting patterns of the $(6 {\times} 4)$ supercell obtained using different exchange-correlation functionals (in combination with appropriate pseudo-potentials). 
The energies (in meV) are given relative to the total energy of the $c(4 {\times} 2)$ reconstruction.
}
\centering
\setlength{\tabcolsep}{5pt}
\begin{tabular}{l|r|r|r}
Tilting           & \multicolumn{1}{c|}{LDA} & \multicolumn{1}{c|}{PBE} & \multicolumn{1}{c}{PBEsol} \\
pattern           & \hphantom{PBEsol}        & \hphantom{PBEsol}        &        \\
\hline\hline
$c(4 {\times} 2)$ &    0.0                   &    0.0                   &    0.0 \\
$p(2 {\times} 2)$ &   38.0                   &   12.3                   &   16.4 \\
SDF               &  105.5                   &  131.4                   &   99.2 \\
TDF               &  103.1                   &  123.9                   &  109.7 \\
SDF-2R            &  241.9                   &  306.8                   &  249.4 \\
$p(4 {\times} 1)$ & 1169.0                   & 1293.6                   & 1220.0 \\
$p(2 {\times} 1)$ &  858.2                   &  942.3                   &  882.4
\end{tabular}
\label{tab:DFTTotalEnergies-v1}
\end{table}


\begin{table*}[t]
\caption{
\textbf{DFT results on interaction energies.}
Interaction energies are given in meV.
For each functional, the left columns are determined from the three energies Eqs.~(\ref{eq:E_c4x2})--(\ref{eq:E_SDF}), with $\Cv{2} = \Cv{3} = \Cd{1} = \Cd{2} = 0$, while the middle columns result from the five energetically lowest structures Eqs.~(\ref{eq:E_c4x2})--(\ref{eq:E_SDF-2R}), with $\Cv{3} = \Cd{2} = 0$.
Finally, the right columns are the result of all structures with the full Hamiltonian Eq.~\eqref{eq:IsingHamiltonian_DFT}.
Note that the resulting renormalized effective NN couplings $J_{\parallel,\perp}$ are equal in all cases.
$\Tc$ is derived by Eq.~\eqref{eq:OnsagerEquation} using the values for $J_{\parallel,\perp}$.
For the derived quantities, we assume an error of $\unit[0.1]{meV}$ in the coupling energies.
The DFT results are compared to the experiment, with the first error bar giving the statistical and the second the systematic error (see also Table~\ref{tab:FitResults}).
}
\centering
\renewcommand{\arraystretch}{1.2}
\setlength{\tabcolsep}{3pt}
\begin{tabular}{c|ddd|ddd|ddd|D{,}{.}{-1}}
           & \multicolumn{3}{c|}{LDA \cite{prb-perdew-81}}
           & \multicolumn{3}{c|}{PBE \cite{prl-perdew-96}}
           & \multicolumn{3}{c|}{PBEsol \cite{prl-perdew-2008}}
           & \multicolumn{1}{c }{\text{Experiment}} \\
Eqs.       & \multicolumn{1}{c }{(\ref{eq:E_c4x2})--(\ref{eq:E_SDF})}
           & \multicolumn{1}{c }{(\ref{eq:E_c4x2})--(\ref{eq:E_SDF-2R})}
           & \multicolumn{1}{c|}{(\ref{eq:E_c4x2})--(\ref{eq:E_p2x1})}
           & \multicolumn{1}{c }{(\ref{eq:E_c4x2})--(\ref{eq:E_SDF})}
           & \multicolumn{1}{c }{(\ref{eq:E_c4x2})--(\ref{eq:E_SDF-2R})}
           & \multicolumn{1}{c|}{(\ref{eq:E_c4x2})--(\ref{eq:E_p2x1})}
           & \multicolumn{1}{c }{(\ref{eq:E_c4x2})--(\ref{eq:E_SDF})}
           & \multicolumn{1}{c }{(\ref{eq:E_c4x2})--(\ref{eq:E_SDF-2R})}
           & \multicolumn{1}{c|}{(\ref{eq:E_c4x2})--(\ref{eq:E_p2x1})}
           & \\
\hline\hline
$\Cv{1}$   & -24.8 & -27.0 & -27.0  & -32.3 & -34.7 & -34.7  & -24.1 & -22.2 & -22.2 &       \\
$\Cv{2}$   &   0   &  -2.2 & -16.6  &   0   &  -2.4 & -14.0  &   0   &   1.9 & -19.3 &       \\
$\Cv{3}$   &   0   &   0   & -14.4  &   0   &   0   & -11.6  &   0   &   0   & -21.3 &       \\
$\Jp$      &       & -24.8 &        &       & -32.3 &        &       & -24.1 &       & -24.9 \pm 0,9 \pm 1.3 \\
\hline
$\Ch{1}$   &  -1.6 &   3.8 &   3.8  &  -0.5 &   5.5 &   5.5  &  -0.7 &   6.4 &   6.4 &       \\
$\Cd{1}$   &   0   &   2.7 &   3.6  &   0   &   3.0 &   3.8  &   0   &   3.5 &   3.7 &       \\
$\Cd{2}$   &   0   &   0   &   0.9  &   0   &   0   &   0.8  &   0   &   0   &   0.2 &       \\
$\Js$      &       &  -1.6 &        &       &  -0.5 &        &       &  -0.7 &       & -0.8 \pm 0,1 \pm 0 \\
\hline
$\Jp/\Js$  
           &       & \multicolumn{1}{D{(}{(}{-1}}{16(1)} &
           &       & \multicolumn{1}{D{(}{(}{-1}}{63(12)} &
           &       & \multicolumn{1}{D{(}{(}{-1}}{35(5)} &
           & 31.2 \pm 3,8 \pm 0 \\
$\Tc$\,(K) 
           &       & \multicolumn{1}{D{(}{(}{-1}}{228(5)} &
           &       & \multicolumn{1}{D{(}{(}{-1}}{210(10)} &
           &       & \multicolumn{1}{D{(}{(}{-1}}{179(7)} &
           & 190.6 \pm 0,4 \pm 9.6 \\
$a_\mathrm{lat}\,(\angstrom)$
           &       &  5.38 &        &       &  5.46 &        &       &  5.43 &       &  5,43
\end{tabular}
\label{tab:DFTCouplingEnergies_NEW}
\end{table*}

The numerical results of our DFT calculations are summarized in Tables~\ref{tab:DFTTotalEnergies-v1} and \ref{tab:DFTCouplingEnergies_NEW} for the functionals used.

\par
From Eqs.~\eqref{eqs:DFTTotalEnergies} we can successively determine numerical values of the interaction energies:
(i)
Firstly, we employed the $c(4 {\times} 2)$, the $p(2 {\times} 2)$ and the single-dimer excitation SDF [see Figs.~\ref{Fig.DFTstructure}(a)--\ref{Fig.DFTstructure}(c)] to uniquely obtain the two effective interaction parameters $\Js$ and $\Jp$ using Eqs.~\eqref{eq:E_c4x2}--\eqref{eq:E_SDF}.
These two couplings will not be altered by the following refinement procedure.
(ii)
Alternatively, one can start by constructing a Hamiltonian focusing only on the five energetically most favorable structures shown in Figs.~\ref{Fig.DFTstructure}(a)--\ref{Fig.DFTstructure}(e).
Using these five structures and Eqs.~\eqref{eq:E_c4x2}--\eqref{eq:E_SDF-2R}, one obtains a Hamiltonian with four interaction parameters, two along the dimer row and two across the dimer row or in the diagonal direction. 
(iii)
Using all seven structures calculated by DFT, we are able to fit a lattice Hamiltonian with six interaction parameters.
The results of these three procedures are summarized in Table~\ref{tab:DFTCouplingEnergies_NEW}.

\par
This procedure provides strong evidence that a two-parameter Hamiltonian with effective interactions is suitable to correctly describe the phase transition on the dimerized Si(001) surface. 
Lattice Hamiltonians including more long-range interactions may still be useful in providing microscopic understanding.
However, due to the alternating sign of the dimer buckling, the more long-ranged interactions, entering with positive and negative signs in Eqs.~\eqref{eqs:Jps}, tend to cancel each other out. 
Thus, it is possible to "condense" these interactions into a compact Hamiltonian with only NN interactions without loss of accuracy.

\par
It is found that the interaction $\Jp < 0$ is strongest along the Si dimer rows, in support of a pronounced anticorrelation between the tilt angles of adjacent dimers along a row. 
Compared to $\Cv{1}$, the NNN interaction $\Cv{2}$ is found to be considerably smaller.
Therefore it seems justified to neglect this interaction in a first analysis.
Considering the even longer ranging NNNN interaction $\Cv{3}$ does not affect the values of $\Cv{1}$, but interestingly $\Cv{2}$ is renormalized.
Still, the sum of both remains the same as well as the value of $\Jp$.
Except for the case when only Eqs.~\eqref{eq:E_c4x2}--\eqref{eq:E_SDF} are considered, the interactions between the dimer rows, $\Ch{1}$, $\Cd{1}$ and $\Cd{2}$, are found to have positive signs, but their combination $\Js$ [Eq.~\eqref{eq:Js}] turns out to be negative in all cases, showing that an alternating tilt angle of the dimers is also preferred across the dimer rows.

\par
Comparing the results obtained with the different DFT functionals, one notes that PBE yields the largest value for the ratio $\Jp/\Js$, while LDA yields the smallest, and PBEsol falls in between.
This trend correlates with the lattice constants predicted by the functionals: While PBE overestimates the lattice constant $a_\mathrm{lat}$ of bulk silicon (DFT value \unit[5.46]{\angstrom} vs.\ experimental value \unit[5.43]{\angstrom}) and LDA underestimates it (\unit[5.38]{\angstrom}), the PBEsol functional gives the most accurate value (\unit[5.43]{\angstrom}).

\par
It has been demonstrated experimentally that the Si(001) surface is under anisotropic surface stress \cite{Men1988}.
The fact that the Si dimers prefer to buckle \emph{alternately} is an indication that the surface attempts to reduce the local surface stress anisotropy \cite{GarciaNorthrup1993}.
Hence, the interaction parameters (that are responsible for alternating buckling on a lattice Hamiltonian) are also expected to be highly sensitive to stress, or equivalently to external strain, as recently demonstrated by means of DFT calculations \cite{Xiao:PRM3.044410}.
We therefore include a discussion of surface stress in the next Sec.~\ref{sec:Stress}.

\par
An intuitive picture of the energetics of the Si(001) surface may be obtained best by looking at the energy differences.
The energy gain for the $c(4 {\times} 2)$ reconstruction over the $p(2 {\times} 2)$ reconstruction obtained in our calculations is \unit[1.4]{meV/dimer} using the PBEsol functional (cf.\ Table~\ref{tab:DFTTotalEnergies-v1}).
In comparison, Ramstad {\it et al.} derived a value of $\unit[(3 \pm 13)]{meV/dimer}$ in their LDA calculations \cite{Ramstad:PRB51.14504}, while in the PBE calculations of Xiao {\it et al.} \cite{Xiao:PRM3.044410} only an energy gain of \unit[0.1]{meV/dimer} was obtained. 
The error bars of these data indicate that such calculations depend on subtle details and choice of the proper functional.
However, irrespective of the DFT functional used, our results confirm that the $c(4 {\times} 2)$ reconstruction is indeed the most stable reconstruction for the dimerized Si(001) surface at $T = \unit[0]{K}$. Moreover, the common trend of the energies reported in Table~\ref{tab:DFTTotalEnergies-v1} among all three functionals give us confidence that DFT gives qualitatively the correct physics of Si dimer buckling.

\par
To locate transition states we used the Climbing-Image Nudged Elastic Band (CI-NEB) \cite{Henkelman:jcp2000} method.
Two different types of processes were considered: the creation of a phason-antiphason pair by flipping one dimer in a perfect $p(2 {\times} 2)$ reconstruction of Si(001), and the barrier for moving an isolated phason by one lattice distance along the Si dimer row.
For the latter case, a $(5 {\times} 2)$ supercell of Si(001) was constructed and the flipping of a single dimer was modelled.
A total of eight images were considered to sample the reaction path of the Si dimer flipping.
Starting from the $p(2 {\times} 2)$ reconstruction, for flipping of a Si dimer the barrier energies of 110, 150, and \unit[74]{meV} are obtained using the LDA, PBE, and PBEsol functionals, respectively.
Since the thermal energy equivalent to the measured (and calculated) $\Tc$ is small compared to the barrier energies, crossing the barrier is a rare event, and thus the thermodynamic modeling with $\sigma_{i,j} = \pm 1$ is justified, as any intermediate tilting angles contribute negligibly to a thermal average.

\par
Finally, we identify the elementary excitations of the $c(4 {\times} 2)$ ground state.
Structurally, the simplest excitation is flipping of a single dimer from its equilibrium tilting position to the opposite tilting.
This situation is depicted as the SDF state in Fig.~\ref{Fig.DFTstructure}.
The excitation energy is given by the energy difference between the SDF state and the ground state and can be calculated from Eq.~\eqref{eq:E_SDF}.
Our calculations yield an excitation energy on the order of \unit[100]{meV} for all three functionals (cf.\ Table~\ref{tab:DFTTotalEnergies-v1}).
The energy barriers for such a transition, i.e., for creating a phason-antiphason pair, are \unit[165]{meV} (LDA), \unit[148]{meV} (PBEsol), and \unit[209]{meV} (PBE), respectively.
It should also be noted that the energy per dimer for the $p(2 {\times} 2)$ reconstruction is lower than for the SDF and TDF states.
However, such non-local collective excitation of the entire surface is unlikely to occur at low temperatures.


\subsection{Surface stress}
\label{sec:Stress}

\par
The surface stress tensor for each tilting pattern of the Si(001) surface is obtained from the $(6 {\times} 4)$ supercell stress tensor as
\begin{align}
    \sigma_{\alpha \beta}^\mathrm{surf} =  c (\sigma_{\alpha \beta}^\mathrm{sc} - \sigma_{zz} \delta_{\alpha \beta} - \tfrac 1 2 \sigma_{\alpha \beta}^\mathrm{sc,H} )\,,
\end{align}
where $c = \unit[26]{\angstrom}$ is the dimension of the supercell normal to the surface, i.e., a thickness of eight atomic layers. 
$\sigma_{\alpha \beta}^\mathrm{sc,H}$ is the supercell surface stress of the Si surface slab passivated on both sides by H atoms.
Including such a term ensures that the contribution of H atoms (at the bottom of the Si slab) to the stress is well excluded.
The presence of $\sigma_{zz}$ in the above expression is a caveat to ensure that a residual nonzero diagonal surface stress component perpendicular to the surface is eliminated.
Thus $\sigma_{\alpha \beta}^\mathrm{surf}$ satisfies the physical requirement of zero $z$ stress \cite{GarciaNorthrup1993}.
The components of the surface stress tensor for each tilting pattern are calculated as $\sigma_{xx}^\mathrm{surf} \mapsto \sigma_\parallel$ and $\sigma_{yy}^\mathrm{surf} \mapsto \sigma_\perp$ for the respective directions along and across the dimer rows (cf.\ Table~\ref{tab:DFTsurfstressAnisotropy}). 
Finally, the stress anisotropy $\Delta \sigma = \sigma_\perp - \sigma_\parallel$ of the tilting patterns is used to quantify the magnitude of the difference between the surface stress tensor components across and along the Si dimer row.
It should be noted that for the $p(4 {\times} 1)$ and $p(2 {\times} 1)$ reconstructions $\sigma_\parallel$ becomes negative, i.e., a compressive stress arises. 
This is in good agreement with experimental \cite{Sato:jphysconmat5.14} and theoretical \cite{vanderbilt:structure.1991} studies, which suggest tensile and compressive stress across and along the dimer rows, respectively, for Si(001) with $p(2 {\times} 1)$ domains formed at $730^\circ$C temperature.
Across the dimer rows the Si surface is subject to tensile stress $\sigma_\perp$, which remains almost the same for each tilting pattern.
This is because it is an individual property of the buckled dimer bond and its length does not change dramatically upon different reconstructions of the Si(001) surface.

\par
Our results show that the stress anisotropy $\Delta \sigma$ becomes smallest (nearly $\unit[39]{meV/\angstrom^2}$ using the PBEsol functional) when Si dimers form the $c(4 {\times} 2)$ and $p(2 {\times} 2)$ tilting patterns.
In contrast, for the tilting patterns $p(4 {\times} 1)$ and $p(2 {\times} 1)$, the surface stress anisotropy becomes very large (104 and $\unit[91]{meV/\angstrom^2}$) because the surface is under compressive stress along the Si dimer row.
Our results are in good agreement with those of previous theoretical \cite{GarciaNorthrup1993, Dabrowski:PRB49.4790} and experimental studies \cite{Yata:PRB74.165407, webb:sursci242.23}.

\par
Looking at the stress anisotropy at different levels of DFT theory gives us clue to correlations between the stress anisotropy and the coupling constant ratio $\Jp/\Js$ with the DFT functionals.
For the $c(4 {\times} 2)$ reconstruction the ratio $\Jp/\Js$ amounts to $16$, $63$, and $35$ for the LDA, PBE and PBEsol functionals, respectively (cf.\ Table~\ref{tab:DFTTotalEnergies-v1}).
From Table~\ref{tab:DFTsurfstressAnisotropy} we derive for the stress anisotropy values of $31$, $64$, and $\unit[39]{meV/\angstrom^2}$ for LDA, PBE and PBEsol functionals, respectively.
Thus, our results show a correlation between the stress anisotropy and the coupling constant ratio $\Jp/\Js$.
As will be shown in the experimental Sec.~\ref{sec:experimental}, the experimental value of this ratio is best reproduced by the PBEsol functional that also closely reproduces the experimental lattice constant of silicon (cf.\ Table~\ref{tab:DFTTotalEnergies-v1}).
Although this is just an observation whose bearing would require more advanced investigations, we note that the Si atoms of the subsurface layer are responsible both for accommodating local strain and for mediating the elastic contribution to the interaction between the Si dimers.
Therefore, we believe that the correlation between the stress anisotropy and the coupling constant ratio is more than just fortuitous.



\begin{table}[t]
\caption{
\textbf{DFT results on surface stress.}
Surface stress $\bar\sigma = (\sigma_\perp + \sigma_\parallel)/2$ and stress anisotropy $\Delta \sigma = \sigma_\perp - \sigma_\parallel$ (in \unit{meV/\angstrom$^2$}) of the tilting patterns of the $(6 {\times} 4)$ supercell obtained using different exchange-correlation levels of theory.
}
\centering
\newcolumntype{e}[1]{D{.}{.}{#1}}
\begin{tabular}{l|e{1}e{1}|e{1}e{1}|e{1}e{1}}
Tilting         & \multicolumn{2}{c|}{LDA}
                & \multicolumn{2}{c|}{PBE}
                & \multicolumn{2}{c }{PBEsol} \\
pattern         & \multicolumn{1}{c }{$\bar\sigma$}
                & \multicolumn{1}{c|}{$\Delta \sigma$}
                & \multicolumn{1}{c }{$\bar\sigma$}
                & \multicolumn{1}{c|}{$\Delta \sigma$}
                & \multicolumn{1}{c }{$\bar\sigma$}
                & \multicolumn{1}{c }{$\Delta \sigma$} \\
\hline\hline
$c(4{\times}2)$ &  65 &  31 &  77 &  64 &  63 &  39 \\
$p(2{\times}2)$ &  62 &  31 &  78 &  65 &  64 &  39 \\
SDF             &  51 &  44 &  69 &  77 &  53 &  53 \\
TDF             &  54 &  42 &  71 &  75 &  55 &  50 \\
SDF-2R          &  41 &  76 &  59 &  94 &  41 &  72 \\
$p(4{\times}1)$ &  33 & 104 &  45 & 128 &  27 & 104 \\
$p(2{\times}1)$ &  38 &  79 &  49 & 116 &  37 &  91
\end{tabular}
\label{tab:DFTsurfstressAnisotropy}
\end{table}

\section{Anisotropic 2D Ising model}
\label{sec:Ising}

In the further analysis of the order-disorder phase transition and of the experimental data, 
we choose to model the system based on the well-known anisotropic 2D Ising model on a rectangular lattice with the Hamiltonian 
\begin{align}
    \mathcal{H}_\mathrm{eff} = - \sum_{i,j} \left( \Jp \sigma_{i,j} \sigma_{i,j+1} + \Js \sigma_{i,j} \sigma_{i+1,j} \right)\,,
\label{eq:IsingHamiltonianJparaJperp}
\end{align}
where $\Jp$ is the effective NN coupling along the dimer row, while $\Js$ is the effective NN coupling between adjacent rows.
NNN couplings can be absorbed additively into the effective NN as in Eqs.~\eqref{eqs:Jps} due to the large correlation length anisotropy \cite{Brand:PRL130.126203}, which guarantees that the corresponding dimers are approximately aligned antiparallel in the whole experimentally relevant temperature range.
This property can be quantified by generalizing the approximation by Zandvliet \cite{Zandvliet:PhaseTrans:2023} to anisotropic systems, yielding a renormalized coupling $\Js = \Ch{1} - [2+\mathcal O(\xip/\xis)^{-2}] \Cd{1}$.

\par
In advantage over more complex models, the 2D Ising model is accessible to an analytical solution \cite{Onsager:PR65.117}, and the well-known scaling laws can be used as a basis for fitting experimental data. 
By analyzing the spot profiles in an electron diffraction experiment along the two directions along and across the dimer rows, the effective NN interaction parameters $J_{\parallel,\perp}$ can be determined experimentally, whereas it is difficult to extract signatures of any more long-ranged interaction parameters from such experimental data.

\begin{figure}
\centering
\includegraphics[width=1.00\columnwidth]{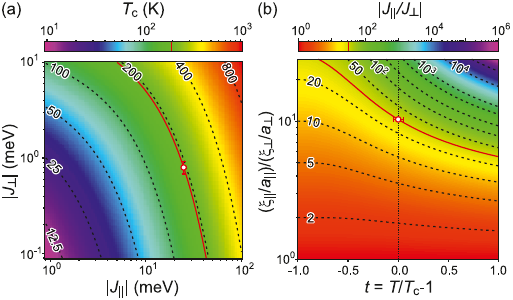}
\caption{
\textbf{Energetics of the anisotropic 2D Ising model.}
(a) Critical temperature $\Tc$ as a function of the effective NN coupling constants $\Jp$ and $\Js$.
(b) Temperature and correlation length ratio dependence of the coupling energy ratio $\lvert \Jp/\Js \rvert$ derived from Eq.~\eqref{eq:CorrelationLength}.
Equation~\eqref{eq:OnsagerEquation} holds at the black-dotted line with $t = 0$ where the relation for $(\xip^+/\alp)/(\xis^+/\als)$ simplifies to Eq.~\eqref{eq:Couplings_from_ratio}.
Red-solid lines and the data points indicate results for $\Tc = \unit[(190.6 \pm 0.4_\mathrm{stat} \pm 9.6_\mathrm{sys})]{K}$ in (a) and $\Jp/\Js = 31.2 \pm 3.8_\mathrm{stat}$ in (b) as derived from the experiment.
}
\label{Fig.Onsager}
\end{figure}

\par
According to the analytical solution of the 2D Ising model by Onsager \cite{Onsager:PR65.117}, the system exhibits a second-order phase transition at the critical temperature $\Tc$ determined by \cite{KramersWannier:PR60.252}
\begin{align}
    \sinh \left( \frac{2 \lvert \Jp \rvert}{\kB \Tc} \right) \, \sinh \left( \frac{2 \lvert \Js \rvert}{\kB \Tc} \right) = 1\,.
\label{eq:OnsagerEquation}
\end{align}
A false-color plot of the solution of Eq.~\eqref{eq:OnsagerEquation} is shown in Fig.~\ref{Fig.Onsager}(a) together with the results of the experiment \cite{Brand:PRL130.126203}.

\par
The spin-spin correlation function $\langle \sigma_{i,j} \sigma_{k,l} \rangle$ decays exponentially on a length scale given by the anisotropic correlation length $\xid$ in direction $\delta \in {\{\parallel,\perp\}}$.
Above (below) $\Tc$, the correlation function asymptotically decays to zero (a constant value); characteristic of a second-order phase transition.   

\par
In the vicinity of the phase transition, however, the system exhibits universal critical behavior, i.e., quantities such as the correlation length $\xid(t)$, the order parameter $\Psi(t)$, or the susceptibility $\chi(t)$ behave asymptotically as power laws of the reduced temperature $t = T/\Tc - 1$,
\begin{subequations}
    \begin{align}
        \Psi(t) &\simeq \Psi^- \left( -t \right)^\beta\,, \label{eq:Psi} \\
        \chi(t) &\simeq \chi^\pm \lvert t \rvert^{-\gamma}\,, \label{eq:chi} \\
        \xid(t) &\simeq \xid^\pm \lvert t \rvert^{-\nu}\,. \label{eq:xi}
    \end{align}
\end{subequations}
Here, $\Psi^-$, $\chi^\pm$ and $\xid^\pm$ are non-universal amplitudes above and below $\Tc$, while the critical exponents have the values $\beta = 1/8$, $\gamma = 7/4$ and $\nu = 1$ for the universality class of the 2D Ising model.
These exponents can accurately describe the behavior of the system close to the critical temperature.
Indeed, $\chi$ generally corresponds to fluctuations that are strong near the critical temperature $\Tc$.

\par
Above $\Tc$, the correlation lengths $\xid$ in the anisotropic 2D Ising model are given by \cite{McCoy+Wu:2D.Ising, HobrechtHucht:SciPostPhys7}
\begin{align}\label{eq:CorrelationLength}
    \frac{\xid(T)}{\ald} & \stackrel{T>\Tc}{=} \left[ \ln \coth \left( \frac{\lvert \Jd \rvert}{\kB T} \right) - \frac{2 \lvert J_{\bar\delta} \rvert}{\kB T}\right]^{-1}\,,
\end{align}
where $\alp = \unit[3.84]{\angstrom}$ is the Si(001) surface lattice parameter, and $\als = 2\alp$, while $\bar\delta$ denotes the direction perpendicular to $\delta$. 
With Eq.~\eqref{eq:OnsagerEquation} and the correlation length ratio $(\xip/\alp)/(\xis/\als)$ [which holds above and below $\Tc$, see Fig.~\ref{Fig.Onsager}(b)] derived from Eq.~\eqref{eq:CorrelationLength}, the renormalized effective NN couplings $\Jd$ of the Si dimers/spins along and across the dimer rows can be derived \cite{Brand:PRL130.126203}:
An expansion of Eq.~\eqref{eq:CorrelationLength} around $\Tc$ from Eq.~\eqref{eq:OnsagerEquation} yields the correlation length amplitudes Eq.~\eqref{eq:xi} above $\Tc$,
\begin{align}\label{eq:CorrelationLengthAmplitudes}
    \frac{\xid^+}{\ald} &= \left[\frac{2 \lvert \Jd \rvert}{\kB\Tc} \sinh\left(\frac{2 \lvert J_{\bar\delta} \rvert}{\kB\Tc}\right) + \frac{2 \lvert J_{\bar\delta} \rvert}{\kB\Tc}\right]^{-1}\,,
\end{align}
from which one can deduce a simple relation between the effective NN couplings $\Jd$ and the ratio of correlation length amplitudes,
\begin{align}
     \frac{\xid^+/\ald}{\xi_{\bar\delta}^+/a_{\bar\delta}} = \sinh\left( \frac{2 \lvert \Jd \rvert}{\kB \Tc} \right)\,,
\label{eq:Couplings_from_ratio}
\end{align}
such that we can determine the anisotropic renormalized effective NN couplings $\Jd$ solely from the correlation length amplitude ratio \cite{LaBella:PRL84.4152}.
Note that the signs of $\Jd$ must be determined from the diffraction analysis below.

\par
In a diffraction experiment, we have access to these quantities by evaluating the spot intensity and its shape.
For $T > \Tc$, the diffraction pattern consists of Lorentzian peaks whose widths $\mathrm{FWHM}_{\mathrm{L},\delta}$ are proportional to the inverse correlation lengths $\xid^{-1}$, while their intensity $I_\mathrm{L}$ is proportional to the susceptibility $\chi$.
For $T < \Tc$, the diffraction pattern consists of Lorentzian peaks just as for $T > \Tc$, but in addition, there are delta peaks whose intensity $I_\mathrm{G}$ scales with the square of the order parameter $\Psi^2$.
Experimentally, all peaks are additionally broadened by the instrumental response function, i.e., a Gaussian function with constant FWHM independent of temperature.


\section{Experiment}
\label{sec:experimental}


\subsection{Experimental methods}
\label{sec:ExpMethods}

\par
We followed the order-disorder phase transition by means of SPA-LEED which combines high resolution in reciprocal space with superior signal-to-noise ratio \cite{Scheithauer:SurfSci178.441, HvH:ZfK214.591}. 
The experiments were performed under ultrahigh vacuum (UHV) conditions at a base pressure $p < \unit[2 {\times} 10^{-10}]{mbar}$ in order to ensure very low surface contamination by residual gas adsorption.
The Si(001) sample (miscut $< 0.2^\circ$) was mounted on a cryostat for sample cooling with liquid nitrogen.
Direct current was applied to heat the sample for degassing at $600^\circ$C and subsequent flash-annealing at $T > 1200^\circ$C for \unit[5]{s} with the pressure remaining in the $\unit[10^{-10}]{mbar}$ regime.
The temperature was monitored during annealing with an IMPAC IGA10 pyrometer.
Subsequently, the sample was cooled to \unit[78]{K} in less than \unit[5]{min}.
Using the built-in resistive heater of the cryostat, the sample was heated from \unit[78]{K} to \unit[400]{K} at a rate of \unit[10]{K/min} while the sample temperature was measured with a Pt100 Ohmic sensor.
The systematic error in temperature determination is in the order of $\pm \unit[10]{K}$ while the statistical error is less than $\pm \unit[1]{K}$.
At the same time, spot profiles through the (00) spot, $p(2 {\times} 1)$ spots and $c(4 {\times} 2)$ spots were continuously taken by SPA-LEED at an electron energy of $E = \unit[112]{eV}$.
The transfer width of our SPA-LEED was larger than \unit[320]{\angstrom}.
The probing electron beam can induce disorder in the $c(4 {\times} 2)$ structure \cite{Matsumoto:PRL90.106103, Shirasawa:PRL94.195502, Mizuno:PRB69.241306, Seino:PRL93.036101}. 
We confirmed that this is not the case in our study because the beam current in SPA-LEED was kept as low as possible.
The total electron dose was smaller by a factor of 50 compared to experiments using back-view LEED optics \cite{Scheithauer:SurfSci178.441}.
Scanning tunneling microscopy (STM) images were taken at room temperature in another chamber under UHV conditions.



\subsection{Experimental results}
\label{sec:ExpResults}

\par
\begin{figure}[t]
\centering
\includegraphics[width=1.000\columnwidth]{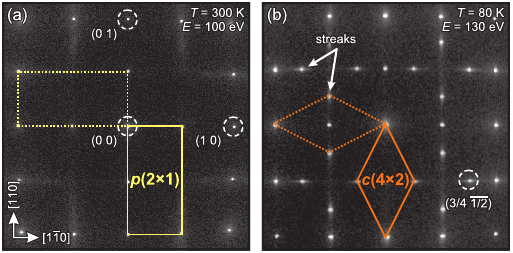}
\caption{
\textbf{SPA-LEED patterns.}
Patterns were taken at (a) $T = \unit[300]{K}$ and (b) $T = \unit[80]{K}$, respectively.
Primitive unit cells of the $p(2 {\times} 1)$ and $c(4 {\times} 2)$ reconstructions are indicated by the yellow rectangles and orange rhombi, respectively.
Additional to the spots streaklike intensity is found centered at the $(3/4~\overline{1/2})$ spots.
}
\label{Fig.Patterns}
\end{figure}

The change of the surface structure upon the Si(001)'s order-disorder phase transition is clearly seen in the LEED patterns shown in Fig.~\ref{Fig.Patterns} and taken at \unit[300]{K} and \unit[80]{K}, i.e., above and below the critical temperature, respectively. 
The diffraction pattern recorded at \unit[300]{K} [Fig.~\ref{Fig.Patterns}(a)] exhibits sharp diffraction spots and a low background reflecting the low step density and low defect and/or adsorbate density.
The pattern is composed of an incoherent superposition of two distinct $p(2 {\times} 1)$ patterns originating from the $90^\circ$-rotated dimer rows arrangement on adjacent terraces.
The twofold periodicity in the diffraction pattern refers to the direction across the dimer rows, while the $\times 1$ periodicity is along the dimer rows [cf.\ Figs.~\ref{Fig.Structure}(a) and \ref{Fig.Patterns}(a)].
From the FWHM (full width at half maximum) of the (00) spot, we determined that the mean terrace width was larger than \unit[50]{nm}, which is in good agreement with the expectation from the miscut of the wafer.

\begin{figure}[t]
\centering
\includegraphics[width=1.000\columnwidth]{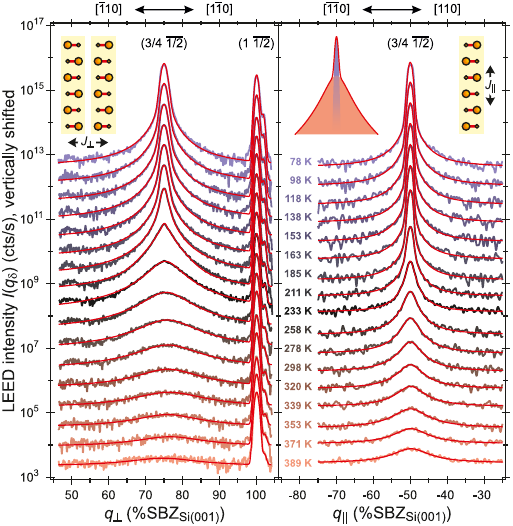}
\caption{
\textbf{Temperature dependence of line profiles through the $(3/4~\overline{1/2})$ spot.}
Profiles through the center of the spot were recorded in the directions across (left panel) and along (right panel) the dimer rows.
Data are plotted as bluish to reddish lines while the fit to the data is given by the solid red lines, see also Eq.~\eqref{eqs:SpotProfile}.
For better visibility the experimental data is Fourier-filtered and 80\% of the linear background is subtracted.
The Gaussian (purple) and Lorentzian (rose) contributions to the spot intensity (corrected for instrumental resolution) are schematically depicted in logarithmic scale closely above $\Tc$ at $T = \unit[198]{K}$ in the right panel.
}
\label{Fig.Lineprofiles}
\end{figure}

The streaklike intensity centered at the quarter-order spot positions is visible even far above the critical temperature \cite{Kubota:PRB49.4810} and is indicative of fluctuations of the dimers, i.e., activation of diffusive phase defects in between $p(2 {\times} 1)$ domains (phasons and antiphasons) \cite{Pennec:PRL96.026102, Hafke:PRL124.016102}.
The faint intensity at the $(1/2~\overline{1/2})$ position (see left panel in Fig.~\ref{Fig.Lineprofiles}) at \unit[50]{\%SBZ} (surface Brillouin zone [$\unit[100]{\%SBZ} = 2\pi/(\unit[3.84]{\angstrom})$]) can originate either from the incoherent overlap of the streaklike intensities of surrounding $c(4 {\times} 2)$ spots and/or from local $p(2 {\times} 2)$ reconstruction sites.
We estimate from the fits to the line profiles shown in Fig.~\ref{Fig.Lineprofiles} that the surface exhibits almost no sites of the latter at any temperature in the investigated range \unit[78-400]{K}.

\par
The diffraction pattern recorded at \unit[80]{K} [Fig.~\ref{Fig.Patterns}(b)] exhibits additional sharp spots at those of the quarter-order positions belonging to a $c(4 {\times} 2)$ reconstruction.
The streaklike intensity is still present.
Again, this pattern is composed of an incoherent superposition of two $90^\circ$-rotated $c(4 {\times} 2)$ patterns. 
Thus, the surface structure has undergone a phase transition to a $c(4 {\times} 2)$ reconstruction with two alternatively buckled dimers per centered unit cell, as sketched in Fig.~\ref{Fig.Structure}(b).

\par
\begin{figure*}[ht]
\centering
\includegraphics[width=1.000\textwidth]{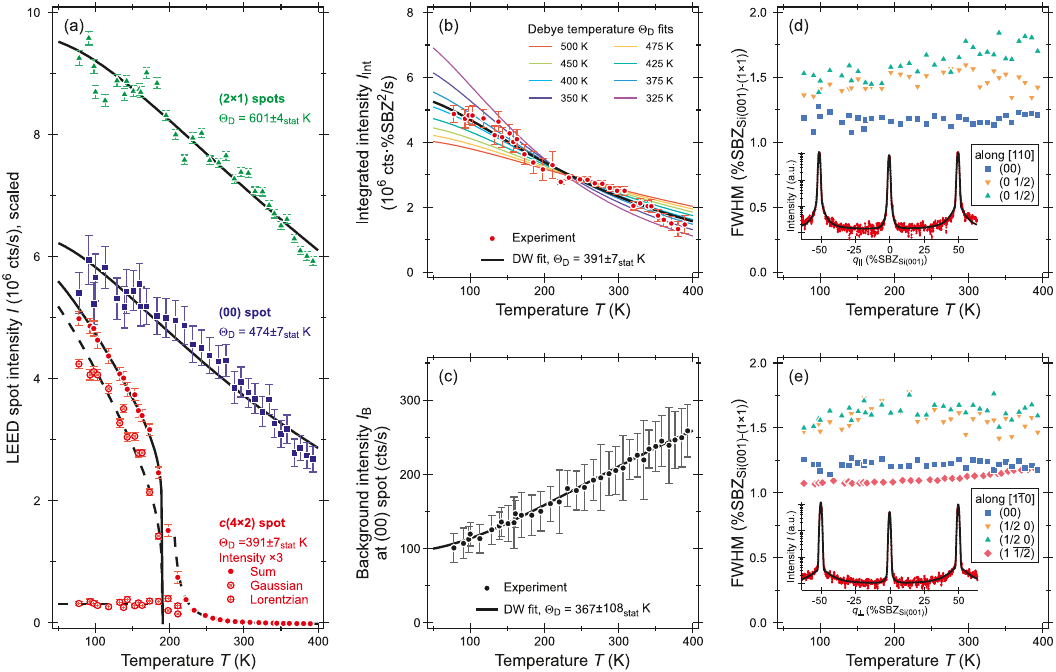}
\caption{
\textbf{Temperature dependence of diffraction spots.}
(a) Scaled LEED spot intensities $I$ of the (00) spot, half-order spots, and the $(3/4~\overline{1/2})$ spot (shown with its Gaussian and Lorentzian contributions).
The spot intensity decreases with increasing temperature due to the Debye-Waller effect according to Eq.~\eqref{eq:DW-Intensity}.
Debye temperatures are indicated.
The $(3/4~\overline{1/2})$ spot clearly deviates from simple Debye-Waller behavior as its intensity is affected by the order-disorder phase transition.
(b) The integrated intensity $I_\mathrm{Int}$ of the $(3/4~\overline{1/2})$ spot drops as function of temperature due to the Debye-Waller effect according to Eq.~\eqref{eq:DW-Intensity} without being affected by the phase transition.
Colored solid lines indicate the expected intensity drop due to the Debye-Waller effect for various Debye temperatures $\TD$ between \unit[325]{K} and \unit[500]{K}.
The black solid line depicts the best fit with $\TD = \unit[(391 \pm 7_\mathrm{stat})]{K}$.
(c) Intensity of the homogeneous thermal diffuse background $I_\mathrm{B}$ and corresponding Debye-Waller fit yielding $\TD = \unit[(367 \pm 108_\mathrm{stat})]{K}$.
(d), (e) FWHMs (uncorrected for instrumental resolution) of the (00) spot and half-order spots along the $\left[ 1 1 0 \right]$ and $\left[ 1 \overline{1} 0 \right]$ direction, respectively.
Insets show line profiles (red dots) and corresponding fits (black lines) through the (00) spot and neighboring half-order spots at $T = \unit[78]{K}$.
}
\label{Fig.Intensities}
\end{figure*}

For a quantitative analysis, spot intensity profiles in the temperature range \unit[78-400]{K} were taken for several diffraction spots.
Line profiles through the (00) spot and the half-order spots exhibit sharp Gaussian-shaped spots reflecting the long-range order of the surface [cf.\ insets of Figs.~\ref{Fig.Intensities}(d) and \ref{Fig.Intensities}(e)].
The FWHM of these spots [see Figs.~\ref{Fig.Intensities}(d) and \ref{Fig.Intensities}(e)] exhibits no temperature dependence, which is also a proof that adsorption of residual gas during this experiment was negligible.
From the sharpest spot, i.e., the $(1~\overline{1/2})$ spot, we determined the instrumental resolution of \unit[1.07]{\%SBZ} [Fig.~\ref{Fig.Intensities}(e)].

\par
The Debye-Waller effect for elastically scattered electrons at temperature $T$ \cite{HvH:ZfK214.591}
\begin{equation}
I(T) = I_0 e^{{-\frac{1}{3}} \langle {\bf u}^{2} \rangle {\bf \Delta K}^2}
\label{eq:Debye-waller}
\end{equation}
for the normalized spot intensity $I(T)/I_0$ with isotropic mean-squared vibrational motion $\langle {\bf u}^2 \rangle$ of the atoms and momentum transfer $\mathbf{\Delta K}$ at almost vertical incidence of $90^\circ - \beta/2 = 86.5^\circ$ ($\beta = 7^\circ$ is the angle between electron gun and channeltron of the instrument \cite{HvH:ZfK214.591}) was employed to disentangle the intensity drop due to the vibrational motion of the atoms and the contributions of the order-disorder phase transition for the peak intensities.
Thus, in the framework of the Debye model, the peak intensity for temperatures below the Debye temperature $\TD$ is given by \cite{Hardy:NIM.86.171}
\begin{subequations}
\label{eq:DW-Intensity}
\begin{align}
    \frac{I(T)}{I_0} &\approx e^{- \frac{3 \hbar^2 \mathbf{\Delta K}^2}{4 m_\mathrm{Si} \kB \TD} \left( 1 + \frac{2}{3} \left( \frac{\pi T}{\TD} \right)^2 \left[ 1 - e^{\left( 1 - \frac{\pi^2}{6}\right) \frac{\TD}{T}} \right] \right) }\,,\label{eq:DW-Intensity_exp}
\intertext{where $m_\mathrm{Si}$ is the mass of the Si atoms, and the momentum transfer in SPA-LEED}
\label{eq:SPA-LEED_Energy}
\lvert \mathbf{\Delta K} \rvert &= \frac{2}{\hbar} \sqrt{ (1 + \cos \beta) m_\mathrm{e} E }\,.
\end{align}
\end{subequations}
Equation~\eqref{eq:DW-Intensity_exp} also considers zero-temperature fluctuations that already become relevant at the low temperatures in our study.

\par
The peak intensities of the relevant spots are plotted in Fig.~\ref{Fig.Intensities}(a).
The intensities of the (00) and half-order spots [blue and green data points in Fig.~\ref{Fig.Intensities}(a)] decrease with increasing temperature, which is explained through a simple Debye-Waller behavior.
Their respective Debye temperatures of $\TD = \unit[(474 \pm 7_\mathrm{stat})]{K}$ for the (00) spot \footnote{A Debye temperature of $\TD = \unit[(488 \pm 11_\mathrm{stat})]{K}$ is determined from the integrated intensity not shown here.} and $\TD = \unit[(601 \pm 4_\mathrm{stat})]{K}$ for the half-order spots are determined using Eq.~\eqref{eq:DW-Intensity} 
[solid-black curves in Fig.~\ref{Fig.Intensities}(a)].
However, the peak intensity of the quarter-order $(3/4~\overline{1/2})$ spot (red data points) clearly deviates from the Debye-Waller behavior as it is strongly affected by the phase transition.
The drop in intensity at $T \approx \unit[200]{K}$ reflects the expected behavior for an order-disorder phase transition.

\par
Figure~\ref{Fig.Intensities}(b) depicts the integrated intensity of the $(3/4~\overline{1/2})$ spot calculated from integration along the line profiles shown below in Fig.~\ref{Fig.Lineprofiles}.
Since the integrated intensity exhibits a small deviation from the expected Debye-Waller behavior around the critical temperature, the fitting of the spot profile may still be affected by the critical behavior of the phase transition, e.g., due to the Fisher exponent $\eta$ \cite{TracyMcCoy:PhysRevB.12.368}.
We have therefore plotted a family of curves describing possible Debye-Waller behaviors using Eq.~\eqref{eq:DW-Intensity} for a range of Debye temperatures from $\TD = \unit[325]{K}$ to \unit[500]{K}.
Analyzing these curves within the kinematic approximation, we can rule out sufficiently strong effects from the critical behavior, and the best fit to the data is obtained for $\TD = \unit[(391 \pm 7_\mathrm{stat})]{K}$, which is in reasonable agreement with the values obtained for the other spots.
In addition, the thermal diffuse background intensity of the diffraction pattern shows an increase with temperature [Fig.~\ref{Fig.Intensities}(c)] as expected from Debye-Waller theory \cite{HvH:ZfK214.591}.
The corresponding Debye-Waller fit yields $\TD = \unit[(367 \pm 108_\mathrm{stat})]{K}$.

\par
The quantitative investigation of the phase transition is employed through a detailed spot profile analysis of the $(3/4~\overline{1/2})$ spot.
Using line profiles, both across (along the $\left[ 1 \overline{1} 0 \right]$ direction, left panel in Fig.~\ref{Fig.Lineprofiles}) and along (along the $\left[ 1 1 0 \right]$ direction, right panel in Fig.~\ref{Fig.Lineprofiles}) the Si dimer rows, the temperature dependence of the spot profiles has been recorded.
The spot profiles consist of a sharp central spike, a broad diffuse part, and background intensity.
In accordance with the 2D Ising model and Refs.~\cite{Kubota:PRB49.4810, Murata:PT53.125}, the data were therefore fitted by a combination of a Gaussian-shaped contribution (sharp central spike) with peak intensity $I_\mathrm{G}(t)$ and full width at half maximum $\mathrm{FWHM}_{\mathrm{G},\delta}$ and a broad Lorentzian-shaped contribution (broad diffuse part) with peak intensity $I_\mathrm{L}(\mathbf{q}_0, t)$ and full width at half maximum $\mathrm{FWHM}_{\mathrm{L},\delta}(t) = 2\pi/\xid(t)$, respectively, where we consider a spot at reciprocal lattice vector $\mathbf{q}_0$, with $\mathbf{q} = (q_\parallel, q_\perp)$.
Such a line profile in direction $\delta$ is therefore described by%
\begin{subequations}\label{eqs:SpotProfile}
\begin{align}
    I(q_\delta,t) &= \delta(q_\delta-q_{0,\delta}) I_\mathrm{G}(t) + I_\mathrm{L}(q_\delta, t)\,,\label{eq:SpotProfile_sum}
\intertext{with asymptotic scaling forms after Debye-Waller correction according to Eq.~\eqref{eq:DW-Intensity}}
    I_\mathrm{G}(t) &\simeq A_\mathrm{G}^- \, (-t)^{2\beta} \,, \label{eq:SpotProfile_G} \\
    I_\mathrm{L}(q_\delta, t) &\simeq A_\mathrm{L}^\pm (y_\delta^\pm)\, |t|^{-\gamma} \,, \label{eq:SpotProfile_L}
\intertext{near $\Tc$, where we introduced the dimensionless scaling variable \cite{TracyMcCoy:PhysRevB.12.368}}
    y_\delta^\pm &= \xid^\pm |t|^{-\nu} \, \frac{q_\delta - q_{0,\delta}}{2\pi} \,. \label{eq:scalingvariable}
\end{align}
\end{subequations}
Here, $I_\mathrm{G}(t)$ describes the sharp central $\delta$-spike proportional to the square of the order parameter $\Psi(t)$ from Eq.~\eqref{eq:Psi} and follows a power law with exponent $2\beta = 1/4$ and amplitude $A_\mathrm{G}^-$ below $\Tc$.
Accordingly, $I_\mathrm{L}(q_\delta,t)$ is the broad diffuse part of the spot profile, whose height is proportional to the susceptibility $\chi(t)$ from Eq.~\eqref{eq:chi} and scales with an exponent of $\gamma = 7/4$, while its width scales anisotropically with the inverse correlation lengths $\xid(t)$ from Eq.~\eqref{eq:xi}.
The corresponding Lorentzian scaling function is denoted $A_\mathrm{L}^\pm(y_\delta^\pm)$ and will be discussed at the end of this section \footnote{Note that in Ref.~\cite[Eq.~(7)]{Brand:PRL130.126203} the factor $|t|^{-\nu}$ is missing in the argument of $A_\mathrm{L}^\pm$.}.

\par
To account for the instrumental response function of the SPA-LEED, a pseudo-Voigtian function (sum of a Lorentzian and a Gaussian peak with the minimum Gaussian FWHM of the sharpest spot, i.e., the $(1~\overline{1/2})$ spot [Fig.~\ref{Fig.Intensities}(e)] as a measure of the instrumental response) was used to fit the Lorentzian contribution \footnote{In principle, to account for the instrumental response function (mainly determined by the shape and size of the channeltron aperture) the profile shape in Eq.~\eqref{eq:SpotProfile_sum} should be convoluted with a purely Gaussian kernel. However, we haven chosen a pseudo-Voigtian instead of a Voigtian function (convolution of Gaussian and Lorentzian) to overcome numerical integrability issues in 2D when evaluating the Lorentzian contribution to the profile. The difference between Voigtian and pseudo-Voigtian profile shape remains rather small in our case.}.
Accordingly, also the Gaussian contribution is broadened by the instrumental response function.
The fit to the data is superimposed on the line profiles in Fig.~\ref{Fig.Lineprofiles} as red curves.

\par
Below $T \approx \unit[200]{K}$ the spot profile of the $(3/4~\overline{1/2})$ spot consists of a sharp Gaussian contribution and a weaker Lorentzian part with constant intensity and FWHM.
Both parts are schematically depicted in the right panel of Fig.~\ref{Fig.Lineprofiles} for $T = \unit[198]{K}$ and as $\otimes$ and $\oplus$ in Fig.~\ref{Fig.Intensities}(a), respectively.
Above $T \approx \unit[200]{K}$ the central spike has disappeared, while the width of the broad diffuse part increases strongly.
However, the diffuse part is still clearly visible at room temperature [see also the streaks of the LEED pattern in Fig.~\ref{Fig.Patterns}(a)] and above.
In contrast, all integer order spots like the (00) spot and all half-order spots always exhibit a spot profile given solely by a sharp Gaussian with constant FWHM independent of temperature [cf.\ Figs.~\ref{Fig.Intensities}(d) and \ref{Fig.Intensities}(e)].

\begin{figure*}[ht]
\centering
\includegraphics[width=1.000\textwidth]{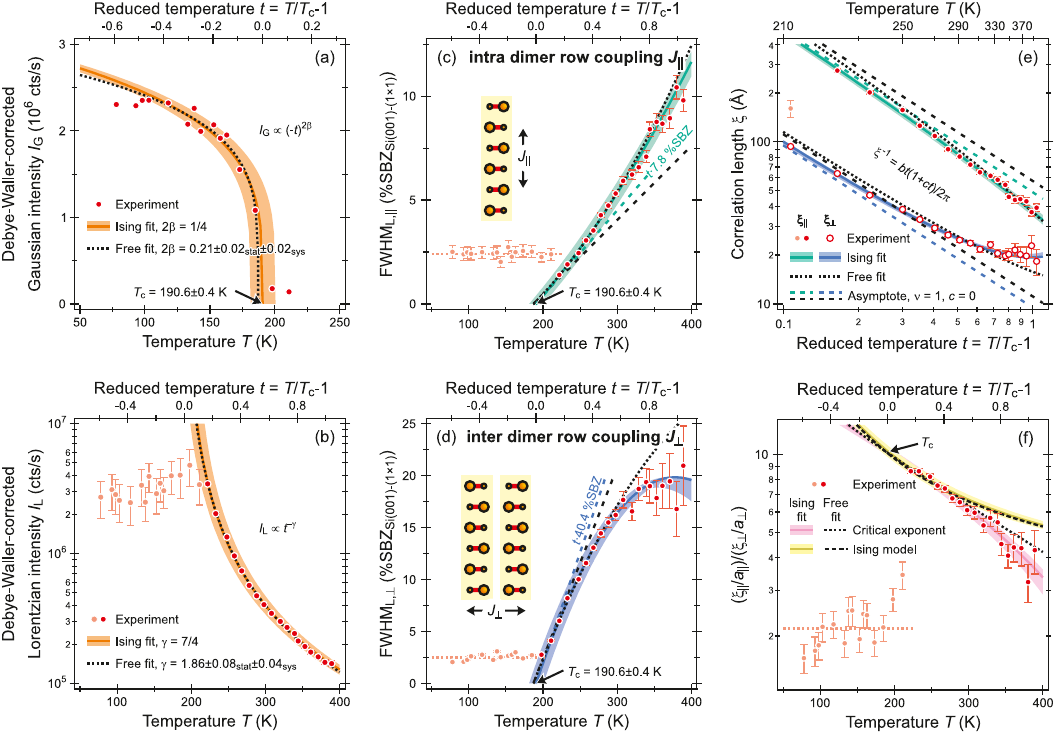}
\caption{
\textbf{Order-disorder phase transition in Si(001).}
Critical behavior of the Gaussian (a) and Lorentzian (b) contributions $I_\mathrm{G,L}$ to the $(3/4~\overline{1/2})$ spot intensity during and above the phase transition, respectively.
Data are corrected for the Debye-Waller effect.
(c) ,(d) Temperature-dependent FWHMs of the Lorentzian contribution to the $(3/4~\overline{1/2})$ spot profile along (along $\left[ 1 1 0 \right]$ direction, intra dimer row coupling) and across (along $\left[ 1 \overline{1} 0 \right]$ direction, inter dimer row coupling) the dimer rows corrected for the instrumental resolution, respectively.
(e) Correlation lengths $\xi_{\parallel,\perp}$ of the Lorentzian contribution vs.\ reduced temperature $t$ above $\Tc$.
(f) Correlation length ratio $(\xip/\alp)/(\xis/\als)$ of the Lorentzian contribution.
Data are fitted for (a) $T \in [\unit[95]{K}, \Tc)$, (b) $T \in [\unit[225]{K}, \unit[400]{K}]$, (c) $T \in [\unit[220]{K}, \unit[300]{K}]$, and (d) $T \in [\unit[210]{K}, \unit[300]{K}]$.
Light-pink data points belong to the so-called domain state and are not taken into account for the fits in (b)--(f).
Colored lines indicate fits to the critical behavior predicted by Onsager theory, determining the critical temperature $\Tc = \unit[(190.6 \pm 0.4_\mathrm{stat} \pm 9.6_\mathrm{sys})]{K}$.
First-order corrections are taken into account for the FWHMs and the correlation lengths in (c)--(e), respectively.
The asymptotic (linear) behavior is shown by the colored-dashed lines in (c)--(e).
In (f) the expected behavior for the 2D Ising model (dark-yellow line) is derived by matching the values with the results from the fit to the critical behavior (dark-pink line) at $\Tc$.
The reduced temperature $t$ is derived from Ising model $\Tc = \unit[190.6]{K}$.
Shaded areas indicate corresponding systematic errors, accordingly.
Black-dotted lines indicate fits to the critical behavior with unfixed exponents $\beta$ and $\gamma$.
}
\label{Fig.Quarter-Spot}
\end{figure*}

\par
The fits to the line profiles of the $(3/4~\overline{1/2})$ spot, namely the Debye-Waller-corrected intensities of the Gaussian central spike $I_\mathrm{G}$ and of the broad diffuse Lorentzian $I_\mathrm{L}$, as well as both Lorentzian peak widths $\mathrm{FWHM}_{\mathrm{L},\delta}$ are further analyzed and compared with the theoretical predictions of the anisotropic 2D Ising model.
Both intensity contributions vary strongly with temperature, i.e., reflecting the critical behavior of the phase transition.
The critical temperature \mbox{$\Tc = \unit[(190.6 \pm 0.4_\mathrm{stat} \pm 9.6_\mathrm{sys})]{K}$} is derived by a global fit of all four critical quantities for temperatures up to \unit[310]{K} [colored-solid lines (shaded areas indicate systematic errors) in Figs.~\ref{Fig.Quarter-Spot}(a)--\ref{Fig.Quarter-Spot}(d)], i.e., $I_\mathrm{G,L}$ and $\mathrm{FWHM}_{\mathrm{L},\delta}$.
The Gaussian contribution [Fig.~\ref{Fig.Quarter-Spot}(a)] exhibits a sharp drop around \unit[200]{K}, indicative of the order-disorder phase transition \cite{Tabata:SurfSci179.L63, Murata:PT53.125}.
The orange-solid line is a fit to the critical behavior following the power law $\left( -t \right)^{2\beta}$ proportional to the square of the order parameter $\Psi$ with $\beta = 1/8$.

\par
For $T > \Tc$ the $(3/4~\overline{1/2})$ spot exhibits a Lorentzian profile with decreasing peak intensity [Fig.~\ref{Fig.Quarter-Spot}(b)] and increasing widths along both directions (cf.\ Fig.~\ref{Fig.Lineprofiles}) as function of temperature.
The orange-solid line is a fit to the intensity above $\Tc$, which is proportional to the susceptibility $\chi$ and scales with the power law $\lvert t \rvert^{-\gamma}$ with $\gamma = 7/4$.

\par
The Lorentzian FWHMs along and across the dimer rows are shown in Figs.~\ref{Fig.Quarter-Spot}(c) and \ref{Fig.Quarter-Spot}(d).
Starting at $\Tc$, both FWHMs increase asymptotically from zero (colored dashed lines) with \mbox{$\mathrm{FWHM}_{\mathrm{L},\parallel}/t = b_\parallel = \unit[(7.8 \pm 0.3_\mathrm{stat} \pm 1.0_\mathrm{sys})]{\%SBZ}$} and $\mathrm{FWHM}_{\mathrm{L},\perp}/t = b_\perp = \unit[(40.4 \pm 0.8_\mathrm{stat} \pm 5.0_\mathrm{sys})]{\%SBZ}$, respectively.
The fit (colored solid lines) to the data shows a deviation from the asymptotes, which is well described up to $T \approx \unit[300]{K}$ by a first-order correction term to the Lorentzian FWHMs, i.e., $\mathrm{FWHM}_{\mathrm{L},\delta}(t) = b_\delta t (1 + c_\delta t + \ldots)$ for $T > \Tc$ \cite{HobrechtHucht:SciPostPhys7}.
Approaching $\Tc$, i.e., $t \to 0^+$, both correlation lengths $\xid(t) = 2\pi/\mathrm{FWHM}_{\mathrm{L},\delta}(t)$ diverge according to Eq.~\eqref{eq:xi} with an exponent $\nu = 1$ [colored-dashed lines in Fig.~\ref{Fig.Quarter-Spot}(e)].

\par
Comparing the experimentally observed temperature dependence of the Lorentzian correlation length ratio $(\xip/\alp)/(\xis/\als)$ above $\Tc$ with the exact solution Eq.~\eqref{eq:CorrelationLength} of the 2D Ising model [see Fig.~\ref{Fig.Quarter-Spot}(f)] we derive by extrapolation $t \to 0^+$,
\begin{align}
\lim_{t\to 0^+}\frac{\xip(t)/\alp}{\xis(t)/\als} = \frac{\als b_\perp}{\alp b_\parallel} = 10.29 \pm 0.39_\mathrm{stat} \pm 0.01_\mathrm{sys}\,.
\label{eq:xi_anisotropy_ratio}
\end{align}
Using Eq.~\eqref{eq:Couplings_from_ratio} at $\Tc$, we finally derive the coupling energies as $\Jp = \unit[(-24.9 \pm 0.9_\mathrm{stat} \pm 1.3_\mathrm{sys})]{meV}$ and $\Js = \unit[(-0.8 \pm 0.1_\mathrm{stat})]{meV}$.
The negative sign of both couplings follows from the spot positions in the diffraction pattern, leading to an antiferromagneticlike coupling of the dimers along and across the dimer rows with $c(4 {\times} 2)$ symmetry.
Eventually, we obtain the coupling energy ratio $\Jp / \Js = 31.2 \pm 3.8_\mathrm{stat}$.

\begin{table}[b]
\caption{\textbf{Critical behavior of the $(3/4~\overline{1/2})$ spot.}
Fit results including statistical (stat, first) and systematic (sys, second) errors.}
\centering
\begin{tabular}{l|r|r}
                                                    & Ising model fit           & Free fit \\
\hline
\hline
$\Tc$ (K)                                           & $190.6 \pm 0.4 \pm 9.6$   & $187.1 \pm 0.6 \pm 9.0$ \\
\hline
$2\beta$                                            & 0.25                      & $0.21 \pm 0.02 \pm 0.02$ \\
$\gamma$                                            & 1.75                      & $1.86 \pm 0.08 \pm 0.04$ \\
\hline
$b_\parallel$ (\%SBZ)                               & $7.8 \pm 0.3 \pm 1.0$     & $6.7 \pm 0.3 \pm 0.3$ \\
$c_\parallel$                                       & $0.32 \pm 0.09 \pm 0.29$  & $0.56 \pm 0.10 \pm 0.09$ \\
$\Jp$ (meV)                                         & $-24.9 \pm 0.9 \pm 1.3$   & $-24.4 \pm 1.2 \pm 1.2$ \\
\hline
$b_\perp$ (\%SBZ)                                   & $40.4 \pm 0.8 \pm 5.0$    & $34.5 \pm 1.0 \pm 1.5$ \\
$c_\perp$                                           & $-0.51 \pm 0.04 \pm 0.22$ & $-0.29 \pm 0.05 \pm 0.06$ \\
$\Js$ (meV)                                         & $-0.8 \pm 0.1 \pm 0$      & $-0.8 \pm 0.1 \pm 0$ \\
\hline
$b_\perp / b_\parallel$                             & $5.15 \pm 0.20 \pm 0$  & $5.16 \pm 0.25 \pm 0$ \\
$\Jp / \Js$                                         & $31.2 \pm 3.8 \pm 0$      & $31.3 \pm 4.9 \pm 0$
\end{tabular}
\label{tab:FitResults}
\end{table}

\par
In addition to the analysis within the anisotropic 2D Ising model, we fitted the critical behavior of the system with free exponents $2\beta$ and $\gamma$.
With this method we can determine how well the Si dimers behave as expected for an Ising system or another universality class.
The results of the global fit are shown in Fig.~\ref{Fig.Quarter-Spot} as black-dotted lines.
For comparison, the results of both fits, i.e., for free exponents and for the anisotropic 2D Ising model with $2\beta = 1/4$ and $\gamma = 7/4$ are summarized in Table~\ref{tab:FitResults}.
As we find for the free fit \mbox{$\Tc = \unit[(187.1 \pm 0.5_\mathrm{stat} \pm 9.0_\mathrm{sys})]{K}$}, \mbox{$2\beta = 0.21 \pm 0.02_\mathrm{stat} \pm 0.02_\mathrm{sys}$}, and $\gamma = 1.86 \pm 0.08_\mathrm{stat} \pm 0.04_\mathrm{sys}$, we can confirm that the system behaves Ising-like.
A similar level of agreement of the experimentally determined exponents and Ising exponents has been found for comparable systems, e.g., the analogous phase transition on Ge(001) \cite{Zandvliet:PhysRep388.1, Cvetko:SurfSci447.L147}.
For the free fit we derive the coupling constants of Si(001) as \mbox{$\Jp = \unit[(-24.4 \pm 1.2_\mathrm{stat} \pm 1.2_\mathrm{sys})]{meV}$} and $\Js = \unit[(-0.8 \pm 0.1_\mathrm{stat})]{meV}$, and thus \mbox{$\Jp / \Js = 31.3 \pm 4.9_\mathrm{stat}$}, very close to the results of the anisotropic 2D Ising model.

\par
The large deviation of the Lorentzian intensity, FWHMs and correlation length ratio for $T \leq \Tc$ (light-pink data points in Fig.~\ref{Fig.Quarter-Spot}) from the expected behavior of the 2D Ising model is due to the formation of finite-sized ordered domains.
The fluctuating dimer system was quenched upon passing the critical temperature during cooling, i.e., reflecting the frozen critical fluctuations at $T > \Tc$.
Here, we observe constant and isotropic Lorentzian FWHMs $\mathrm{FWHM}_{\mathrm{L},\perp} = \unit[(2.52 \pm 0.01_\mathrm{stat})]{\%SBZ}$ and $\mathrm{FWHM}_{\mathrm{L},\parallel} = \unit[(2.41 \pm 0.07_\mathrm{stat})]{\%SBZ}$, respectively, i.e., the resulting domain size is limited by the quench to about \unit[14]{nm}.

\begin{figure}[t]
    \centering
    \includegraphics[width=\columnwidth]{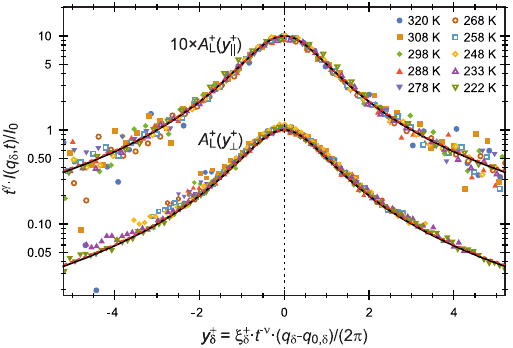}
    \caption{Scaling plot of the Lorentzian scaling function $A_\mathrm{L}^+(y_\delta^+)$ from Eq.~\eqref{eq:SpotProfile_L} above $\Tc$, with direction $\delta\in\{\parallel,\perp\}$ and scaling variable $y_\delta^+$ from Eq.~\eqref{eq:scalingvariable}. 
    The black-solid line is the Ornstein-Zernike prediction $A_\mathrm{OZ}(y) = (1+y^2)^{-1}$, while the red-dashed line is the exact scaling function from Ref.~\cite[Eq.~(B7)]{TracyMcCoy:PhysRevB.12.368}. \label{fig:scalingplot}
    }
\end{figure}

\par
Finally, we analyze the Lorentzian scaling function $A_\mathrm{L}^+(y_\delta^+)$ above $\Tc$ from Eq.~\eqref{eq:SpotProfile_L}, with direction \mbox{$\delta \in \{\parallel,\perp\}$} and scaling variable $y_\delta^+$ from Eq.~\eqref{eq:scalingvariable}. 
For this, in Fig.~\ref{fig:scalingplot} we plot the Debye-Waller-corrected experimental data for temperatures $\unit[222]{K} \leq T \leq \unit[320]{K}$ as a scaling plot, by rescaling the abscissa $q_\delta-q_{0,\delta}$ with the correlation length $\xid(t)$ from Eq.~\eqref{eq:xi} and the ordinate $I_\mathrm{L}(q_\delta, t)$ with the inverse susceptibility $\chi(t)^{-1}$ from Eq.~\eqref{eq:chi}.
We obtain an impressive quantitative data collapse of the different intensities in both directions without applying any further fitting parameters. 
The systematic deviations at negative $y_\perp^+$ are due to the enhanced intensity at $q_\perp = \unit[50]{\%SBZ}$ originating from the neighboring quarter-order spots (cf.\ left panel of Fig.~\ref{Fig.Lineprofiles}).
The data fall nicely onto the Ornstein-Zernike prediction $A_\mathrm{OZ}(y) = (1+y^2)^{-1}$ (black-solid curve). 
We also show the exact scaling function obtained by Tracy and McCoy \cite[Eq.~(B7)]{TracyMcCoy:PhysRevB.12.368} (red dashed curve), which is barely distinguishable from $A_\mathrm{OZ}(y)$ at these small values of $|y|$, and only starts to deviate for $|y|\gtrsim 10$, where it decays as $|y|^{-7/4}$ instead of $y^{-2}$.
This close agreement does not allow us to discuss deviations from Ornstein-Zernike scaling, as has been done for other surface Ising systems \cite{Campuzano:PRL54.2684}, in the temperature and wavevector range covered by our present experiment.


\section{Discussion \& conclusions}
\label{sec:Disc+Concl}

\par
Comparing the interaction parameters derived from the electron diffraction experiment via fitting to the Ising model, $\Jp = \unit[(-24.9 \pm 0.9_\mathrm{stat} \pm 1.5_\mathrm{sys})]{meV}$ and $\Js = \unit[(-0.8 \pm 0.1_\mathrm{stat} \pm 0.2_\mathrm{sys})]{meV}$ with \mbox{$\Tc = \unit[(190.6 \pm 0.4_\mathrm{stat} \pm 9.6_\mathrm{sys})]{K}$} with the values calculated from DFT (Table~\ref{tab:DFTCouplingEnergies_NEW}), we conclude that all the three considered functionals predict the critical temperature in the correct range, \unit[179(7)-228(5)]{K}, with the PBEsol functional coming closest to the experimental value of $\Tc$.
However, the anisotropy ratio $\Jp/\Js$ varies strongly, between $16(1)$ for the LDA and $63(12)$ for the PBE functional.
Again, the PBEsol functional, which yields an anisotropy ratio of $35(5)$, comes closest to the experimental value of $31.2 \pm 3.8_\mathrm{stat}$ in the Ising model fit ($31.3 \pm 4.9_\mathrm{stat}$ in the free fit).
This demonstrates a compensation effect: a functional that overestimates $\Jp$ tends to underestimate $\Js$ and vice versa.
Since these two interactions enter in opposite ways into Eq.~\eqref{eq:OnsagerEquation} for $\Tc$, the predicted value of $\Tc$ is quite robust.
Only with the help of subtle experimental methods, i.e., the spot profile analysis of diffraction spots, has the anisotropy parameter become experimentally accessible, allowing us to evaluate the performance of the different DFT functionals.
We find that the PBEsol functional, whose construction was motivated by improving the values of surface energies, also performs best in predicting the critical behavior of the order-disorder phase transition.
We trace back this excellent performance of the PBEsol functional to yielding the correct lattice constant and the surface stress anisotropy associated with the Si(001) surface reconstruction.
The PBEsol results, yielding $\Jp = \unit[-24.1]{meV}$, $\Js = \unit[-0.7]{meV}$, and $\Tc = \unit[179(7)]{K}$, are strikingly close to the experiment, within overlapping error bars.

\par
Discrepancies between the experimental results and earlier theoretical predictions of $\Tc$ have been interpreted as a result of overestimated defect concentrations in the percent range within the dimer structure \cite{Inoue:PRB49.14774, Saxena:SurfSci160.618, Nakamura:PRB52.8231, Nakamura:PRB55.10549, Okamoto:PRB62.12927, Osanai:SurfSci493.319, Natori:ASS212-213.705}.
We have shown in this study that such an explanation is incorrect for clean surfaces with low defect concentrations.
The critical temperature of $\Tc = \unit[190.6]{K}$ is a robust result because it is directly related to the interaction energies of the dimers via Eq.~\eqref{eq:OnsagerEquation}.

\par
The very small value of $\Js$ and the pertinent small energy difference between the $c(4 {\times} 2)$ and $p(2 {\times} 2)$ structure are also consistent with the coexistence of $c(4 {\times} 2)$ and $p(2 {\times} 2)$ domains sometimes observed in STM images at low temperatures \cite{Shigekawa:JJAP36.L294}.
Even slight perturbations of the surface by the electric field and/or the current generated by the STM tip may be sufficient to switch between the two reconstructions.
In addition, low-temperature STM experiments at elevated tip voltage show dynamic changes of the Si dimer buckling \cite{Pennec:PRL96.026102, Sagisaka:PRL91.146103}, in particular at the boundary between the $c(4 {\times} 2)$ and $p(2 {\times} 2)$ domains.
The (mobile) surface excitations observed in such an experiment are called phasons \cite{Shigekawa:JJAP36.L294}, where a phason (or antiphason) corresponds to two adjacent Si dimers tilted in the same direction in an otherwise completely ordered environment.
Phasons can move along a Si dimer row and can only be created or annihilated as phason-antiphason pairs.
In this sense, they are topological defects in an ordered 1D structure.
A Si dimer chain bounded by a phason and an antiphason can be viewed as a domain whose order is out-of-phase with the surrounding $c(4 {\times} 2)$ structure.
The finite length of such a dimer chain leads to a broadening of the $(3/4~\overline{1/2})$ diffraction spot in a LEED image.
Therefore the Lorentzian FWHM of this quarter-order LEED spot is a measure of the phason density.
Thus, the size of ordered domains is determined by a geometric length distribution \cite{Wollschlaeger:SurfSci328.325}, i.e., the correlations decay on large length scales $r$ for $T > \Tc$ as $\propto \exp(-r/\xid)$ \cite{McCoy+Wu:2D.Ising}.


\section{Summary \& outlook}
\label{sec:Summ+Outl}

\par
Our comprehensive DFT and SPA-LEED results provide microscopic understanding of second-order phase transitions on surfaces.
The dimerized Si(001) surface is a prototypical system that exhibits such a phase transition at a critical temperature of $\Tc = \unit[190.6]{K}$.
Due to the dimerization of the topmost surface layer on the square bulk-terminated Si(001) face, the dimers are arranged in dimer rows and thus anisotropic coupling between the dimers is observed.
Mapping of the anisotropic 2D Ising model for spins on a rectangular lattice onto the structure of Si dimers allows for a quantitative theoretical description of the experimentally observed phase transition.
The microscopic details of the (electrostatic) coupling between the individual dimers described by the Ising model determine the critical temperature of the order-disorder phase transition.
Our DFT calculations go a step beyond the simple anisotropic 2D Ising model by considering longer-ranging interactions.
At the same time, the effective lattice Hamiltonian derived from our DFT results yields the same critical behavior as the anisotropic 2D Ising model.

\par
The understanding of the order-disorder phase transition gained in the framework of the anisotropic 2D Ising model allows us to analyze non-equilibrium states of the Si(001) surface that could be obtained, e.g., by rapid cooling of the surface below $\Tc$.
If the cooling is sufficiently fast, different domains of the $c(4 {\times} 2)$ structure, bounded by a phason and antiphason, could be frozen in, analogous to the kinetically limited domain structure already observed in our experiments, as reported in Sec.~\ref{sec:experimental}.
This could be the subject of future research, e.g., in experiments where the cooling rate is systematically varied over a wide range. 

\par
From the theoretical perspective, general considerations concerning the creation and persistence of defects while swiftly crossing a phase transition can be helpful in making predictions about the domain structure.
The density of phasons and antiphasons can only be reduced, i.e., a perfectly ordered surface can only be achieved, if these defects have sufficient time to propagate, meet and annihilate.
Thus, small domains can be eliminated during cooling, while larger ones will persist even below $\Tc$.
Given that the Kibble-Zurek mechanism \cite{Kibble:JoPA9.1387, Zurek:Nature317.505} applies, a scaling law is predicted: the initial size of the domains that persist during cooling scales like $(\tau_Q/\tau_0)^\alpha$, where $\tau_Q$ is the time scale associated with cooling, defined by $t/\tau_Q = 1 - T(t)/\Tc$, and $\tau_0$ is an intrinsic time scale of the system, here associated with the flipping rate of the Si dimers \cite{Schaller:arXiv.2310.18216}.
The exponent $\alpha \approx 0.3$ is related to the dynamic scaling exponent $z$ of the underlying 2D Ising model.
Thus, studying the Si(001) surface under non-equilibrium conditions in combination with a scaling analysis could provide valuable information about the microscopic time scale associated with Si dimer flipping.


\appendix*
\section{Comparison to prior lattice Hamiltonians}
\label{sec:Appendix}

\par
In early studies \cite{Ihm:PRL51.1872, Alerhand:PRB35.5533, Inoue:PRB49.14774}, researchers included in their Hamiltonians three interactions $\CV$, $\CH$, and $\CD$, along, across, and diagonal to the Si dimer rows (cf.\ Table~\ref{tab:DFTCouplingEnergies_Comprehensive_app}).  
Later, conditional longer-ranged interactions ($\CViii$ and $\CViiii$) were added \cite{Pillay:SurfSci554.150, Xiao:PRM3.044410}, which contribute only when three or four adjacent dimers are tilted in the same direction. 
In this paper, we explore two approaches, simply adding $\CViii$ or adding both $\CViii$ and $\CViiii$.
To retain consistency with the literature on spin models, we prefer an equivalent representation of the Hamiltonians in terms of (unconditional) NNN interactions $\CViii$ and quadruple interactions $\CViiii$. 
Thus, the generalized Ising model Hamiltonians used to fit our DFT results read 
\begin{align}
\label{eq:IsingHamiltonian_DFT1_app}
\mathcal{H}_0
 = {} & - \CV \sum_{i,j} \sigma_{i,j} \sigma_{i,j+1}
- \CH \sum_{i,j} \sigma_{i,j} \sigma_{i+1,j} \nonumber \\
& - \CD \sum_{i,j} \sigma_{i,j} ( \sigma_{i+1,j+1} + \sigma_{i+1,j-1} ) \nonumber \\
& - \CViii \sum_{i,j} \Big[ 1 + \sum_{0\leq r < s < 3} \sigma_{i,j+r} \sigma_{i,j+s} \Big]\,, \\
\mathcal{H}_1 
 = {}&  \mathcal{H}_0 - \CViiii \sum_{i,j} \Big[ 1 
    + \sigma_{i,j} \sigma_{i,j+1} \sigma_{i,j+2} \sigma_{i,j+3} \nonumber\\
& \hphantom{{}-\CViii \sum_{i,j} \Big[ 1} + \sum_{0\leq r < s < 4} \sigma_{i,j+r} \sigma_{i,j+s} \Big]
  \,.
\label{eq:IsingHamiltonian_DFT2_app}
\end{align}

\par
The coupling parameters of the lattice Hamiltonians given in Eqs.~\eqref{eq:IsingHamiltonian_DFT1_app} and \eqref{eq:IsingHamiltonian_DFT2_app} are determined by solving a system of linear equations corresponding to the tilting patterns considered in Fig.~\ref{Fig.DFTstructure}.
After subtracting the energy of the reference ground state $c(4 {\times} 2)$, we find
\begin{subequations}
\begin{align}
E_{c(4 {\times} 2)} & = 0 \,, \label{eq:E_c4x2_app}\\
E_{p(2 {\times} 2)} & = -24(  \CH       - 2 \CD) \,, \label{eq:E_p2x2_app}\\
E_\mathrm{SDF}      & = -4 (  \CH + \CV - 2 \CD +  \CViii) \,, \label{eq:E_SDF_app}\\
E_\mathrm{TDF}      & = -4 (2 \CH + \CV - 4 \CD)           \,, \label{eq:E_TDF_app}\\
E_\mathrm{SDF-2R}   & = -8 (        \CV - 2 \CD +  \CViii) \,, \label{eq:E_SDF-2R_app}\\
E_{p(4 {\times} 1)} & = -24(        \CV - 2 \CD + 2\CViii + 4\CViiii)\,, \label{eq:E_p4x1_app}\\
E_{p(2 {\times} 1)} & = -24(  \CH + \CV         + 2\CViii + 4\CViiii)\,. \label{eq:E_p2x1_app}
\end{align}
\label{eqs:DFTTotalEnergies_app_new}
\end{subequations}
To determine the interaction parameters of $\mathcal{H}_0$ we considered only the five energetically lowest structures (cf.\ Table~\ref{tab:DFTTotalEnergies-v1}), Eqs.~\eqref{eq:E_c4x2_app}--\eqref{eq:E_SDF-2R_app}.

\begin{table}[ht]
\caption{
\textbf{Comparison to previous results.}
Interaction energies are given in meV.
$\Tc$ is derived by Eq.~\eqref{eq:OnsagerEquation} using the values of $\Jd$.
Energies $\CViii$ and $\CViiii$ from literature are converted to our re-definition according to Eqs.~\eqref{eq:IsingHamiltonian_DFT1_app} and \eqref{eq:IsingHamiltonian_DFT2_app}.
The * indicates a phase transition from the disordered $p(2 {\times} 1)$ 
phase to the $p(2 {\times} 2)$ phase as ground state instead of the $c(4 {\times} 2)$ phase.
Note that Xiao \textit{et al.} \cite{Xiao:PRM3.044410}, when quoting their numbers for $\CH$ and $\CD$ in their Table~II, forgot to divide by a factor of 2.
This must be concluded from their total energies in Table~I that are used as input to determine $\CH$ and $\CD$.
Also their $\CViiii$ was miscalculated.
}
\centering
\begin{tabular}{l|rr|r|r|rr}
    & \multicolumn{2}{c|}{LDA}   & \multicolumn{1}{c|}{PBE} & \multicolumn{1}{c|}{PW91}   & \multicolumn{2}{c}{TB} \\ 
\hline
\hline
    & \multicolumn{1}{c}{Inoue} & \multicolumn{1}{c|}{Pillay} & Xiao & Pillay & \multicolumn{1}{c}{Ihm} & \multicolumn{1}{c}{Fu} \\
    & \multicolumn{1}{c}{\cite{Inoue:PRB49.14774}} & \multicolumn{1}{c|}{\cite{Pillay:SurfSci554.150}} & \multicolumn{1}{c|}{\cite{Xiao:PRM3.044410}} & \multicolumn{1}{c|}{\cite{Pillay:SurfSci554.150}} & \multicolumn{1}{c}{\cite{Ihm:PRL51.1872}} & \multicolumn{1}{c}{\cite{Fu:SurfSci494.119}}  \\
\hline 
$\CV = \Jp$         & $-51.9$   & $-16.8$   & $-37.4$   & $-25.3$   & $-26.0$   & $-32.4$   \\
$\CH$               & 6.6       & 6.4       & 7.9       & 7.5       & 10.0      & 0.6       \\
$\CD$               & 3.6       & 4.2       & 4.0       & 4.1       & 4.0       & 0.6       \\
$\CViii$            & --        & 0.1       & $-3.7$    & $-4.2$    & --        & $-2.7$    \\
$\CViiii$           & --        & $-5.3$    & $-1.9$    & $-2.7$    & --        & --        \\
$\Js = \CH - 2\CD$  & $-0.6$    & $-2.0$    & $-0.1$    & $-0.7$    & 2.0       & $-0.6$    \\
\hline
$\Jp / \Js$         & 86.5      & 8.4       & 374.0     & 36.7      & $-13.0$*  & 54.0      \\
\hline
$\Tc$ (K)           & 315.8     & 186.6     & 173.4     & 187.1     & 252.5     & 218.3
\end{tabular}
\label{tab:DFTCouplingEnergies_Comprehensive_app}
\end{table}

\par
For comparison with previous literature (cf.\ Table~\ref{tab:DFTCouplingEnergies_Comprehensive_app}), we have also parameterized a Hamiltonian $\mathcal{H}_1$ [Eq.~\eqref{eq:IsingHamiltonian_DFT2_app}] and included the resulting interaction parameters in Table~\ref{tab:DFTCouplingEnergies_app}.
The equated results to $\mathcal{H}_1$ are based on all equations except Eq.~\eqref{eq:E_SDF-2R_app}, as this was the previously used set of patterns.
With the exception of the early results of Inoue {\it et al.} \cite{Inoue:PRB49.14774} who obtained an unusually large value of $\CV$, the results are in reasonable agreement with each other. 
In particular, the tight-binding (TB) calculations also describe the leading interaction term $\CV$ with the correct magnitude. 
However, neglecting both $\CViii$ and $\CViiii$ (cf.\ Refs.~\cite{Inoue:PRB49.14774, Ihm:PRL51.1872}) leads to an overestimation of the remaining coupling parameters.

\begin{table}[ht]
\caption{
\textbf{DFT results on interaction energies II.}
Interaction energies are given in meV.
$\Tc$ is derived by Eq.~\eqref{eq:OnsagerEquation} using the values of $\Jd$.
Note the flipped sign of $\CViii$ for PBEsol exchange-correlation functional.
}
\centering
\begin{tabular}{l|rr|rr|rr}
                            & \multicolumn{2}{c|}{LDA}  & \multicolumn{2}{c|}{PBE}  & \multicolumn{2}{c}{PBEsol}  \\ 
\hline
\hline
$\CV = \Jp$                 & $-22.6$   & $-22.6$       & $-29.9$   & $-29.9$       & $-26.1$   & $-26.1$         \\
$\CH$                       & 3.8       & 5.7           & 5.5       & 7.1           & 6.4       & 6.7             \\
$\CD$                       & 2.7       & 3.6           & 3.0       & 3.8           & 3.5       & 3.7             \\
$\CViii$                    & $-2.2$    & $-2.2$        & $-2.4$    & $-2.4$        & 1.9       & 1.9             \\
$\CViiii$                   & --        & $-3.6$        & --        & $-2.9$        & --        & $-5.3$          \\
$\Js = \CH - 2\CD$          & $-1.6$    & $-1.6$        & $-0.5$    & $-0.5$        & $-0.7$    & $-0.7$          \\
\hline
$\Jp / \Js$                 & 14.3      & 14.3          & 58.4      & 58.4          & 38.2      & 38.2           \\
\hline
$\Tc$ (K)                   & 213.6     & 213.6         & 198.2     & 198.2         & 190.2     & 190.2          \\
\hline
$a_{\rm lat} (\angstrom)$   & \multicolumn{2}{c|}{5.38}  & \multicolumn{2}{c|}{5.46}  & \multicolumn{2}{c}{5.43}
\end{tabular}
\label{tab:DFTCouplingEnergies_app}
\end{table}

\par
It should be noted that the selection of tilting patterns for the computation of the interaction energies of the Hamiltonian $\mathcal{H}_0$ (cf.\ Table~\ref{tab:DFTCouplingEnergies_app}), in particular the $p(2 {\times} 1)$ and $p(4 {\times} 1)$ structures, leads to much larger values of $\CH$ and $\CD$ compared to the Hamiltonian $\mathcal{H}_0$.
However, these two interactions of opposite sign largely cancel out when the effective interaction $\Js$ is calculated. 
Moreover, the inclusion of the energetically high-lying $p(2 {\times} 1)$ structure in the calculation forces an unphysically large value of $\CViiii$.
To date, neither the $p(2 {\times} 1)$ nor the $p(4 {\times} 1)$ structure have been observed experimentally as long-range ordered reconstructions. 
This is consistent with our finding that both patterns are thermodynamically unfavorable by $\approx \unit[0.85-1.3]{eV}$ compared to the other five patterns considered over all exchange-correlation functionals used in the present paper.
In contrast, the four interaction energies $\CV$, $\CH$, $\CD$, and $\CViii$ are determined using the five most stable tilting patterns.
Thus, we consider the Hamiltonian $\mathcal{H}_0$ with four interaction terms to be physically more realistic, and thus better describing the underlying physics of the Si(001) surface than the Hamiltonian $\mathcal{H}_1$ with five interaction terms.
In fact, as we showed in Sec.~\ref{sec:DFT-results}, even two {\em effective} interaction parameters, $\Jp$ and $\Js$, are sufficient to describe the critical behavior close to the order-disorder phase transition.

\par
We have also carried out the surface stress tensor calculation for a thicker slab (10 Si atomic layers) and we realized that, except for the $p(2{\times}1)$ tilting pattern, the components of the surface stress along and across the Si dimer rows change just a little, see Table~\ref{tab:DFTsurfstressAnisotropy-revsion}.
However, the stress anisotropy undergoes only very slight changes since the stress components vary almost equally by increasing the number of layers.

\begin{table}
\caption{
\textbf{DFT results on surface stress.}
Surface stress $\sigma_{\parallel,\perp}$ (in \unit{meV/\angstrom$^2$}) of the tilting patterns of the $(6 {\times} 4)$ supercell as well as stress anisotropy $\Delta \sigma$ obtained using PBEsol exchange-correlation functional for different thicknesses (8 and 10 atomic layers) of the Si(001) surface slab.
Negative (positive) signs denote compressive (tensile) stress.
}
\centering
\newcolumntype{e}[1]{D{.}{.}{4,1}}
\begin{tabular}{l|e{1}e{1}e{1}|e{1}e{1}e{1}|e{1}e{1}e{1}}
Tilting         & \multicolumn{3}{c|}{8 layers}
                & \multicolumn{3}{c }{10 layers}
\\
pattern         & \multicolumn{1}{c }{$\sigma_\perp$}
                & \multicolumn{1}{c }{$\sigma_\parallel$}
                & \multicolumn{1}{c|}{$\Delta \sigma$}
                & \multicolumn{1}{c }{$\sigma_\perp$}
                & \multicolumn{1}{c }{$\sigma_\parallel$}
                & \multicolumn{1}{c }{$\Delta \sigma$}
 \\
\hline\hline
$c(4{\times}2)$ & \multicolumn{1}{e{1} }{82.4}
                & \multicolumn{1}{e{1} }{43.3}
                & \multicolumn{1}{e{1}|}{39.1}
                & \multicolumn{1}{e{1} }{79.5}
                & \multicolumn{1}{e{1} }{40.4}
                & \multicolumn{1}{e{1} }{39.2}
\\
$p(2{\times}2)$ & \multicolumn{1}{e{1} }{83.5}
                & \multicolumn{1}{e{1} }{45.0}
                & \multicolumn{1}{e{1}|}{38.6}
                & \multicolumn{1}{e{1} }{81.2}
                & \multicolumn{1}{e{1} }{43.0}
                & \multicolumn{1}{e{1} }{38.2}
\\
SDF             & \multicolumn{1}{e{1} }{78.9}
                & \multicolumn{1}{e{1} }{26.2}
                & \multicolumn{1}{e{1}|}{52.7}
                & \multicolumn{1}{e{1} }{78.0}
                & \multicolumn{1}{e{1} }{24.8}
                & \multicolumn{1}{e{1} }{53.2}
\\
TDF             & \multicolumn{1}{e{1} }{80.1}
                & \multicolumn{1}{e{1} }{30.3}
                & \multicolumn{1}{e{1}|}{49.8}
                & \multicolumn{1}{e{1} }{78.5}
                & \multicolumn{1}{e{1} }{29.0}
                & \multicolumn{1}{e{1} }{49.5}
\\
SDF-2R          & \multicolumn{1}{e{1} }{76.6}
                & \multicolumn{1}{e{1} }{5.0}
                & \multicolumn{1}{e{1}|}{71.6}
                & \multicolumn{1}{e{1} }{74.7}
                & \multicolumn{1}{e{1} }{4.3}
                & \multicolumn{1}{e{1} }{70.3}
\\
$p(4{\times}1)$ & \multicolumn{1}{e{1} }{78.9}
                & \multicolumn{1}{e{1} }{-24.9}
                & \multicolumn{1}{e{1}|}{103.9}
                & \multicolumn{1}{e{1} }{78.6}
                & \multicolumn{1}{e{1} }{-26.0}
                & \multicolumn{1}{e{1} }{104.6}
\\
$p(2{\times}1)$ & \multicolumn{1}{e{1} }{82.2}
                & \multicolumn{1}{e{1} }{-8.5}
                & \multicolumn{1}{e{1}|}{90.7}
                & \multicolumn{1}{e{1} }{80.2}
                & \multicolumn{1}{e{1} }{-12.9}
                & \multicolumn{1}{e{1} }{93.1}
\end{tabular}
\label{tab:DFTsurfstressAnisotropy-revsion}
\end{table}


\section*{Author contributions}

\par
G.J.\ and R.H.\ performed the experiments.
H.M.\ performed DFT calculations.
C.B.\ and A.H.\ analyzed the data and prepared the figures.
C.B., H.M., A.H., P.K., and M.H.-v.H.\ drafted the manuscript.
M.H.-v.H., A.H., B.S., P.K., and R.S.\ conceived and supervised the project.
The manuscript was written through contributions of all authors.
All authors have given approval to the final version of the manuscript.

\par
The authors declare no competing financial interest.


\section*{Acknowledgements}

\par
Funded by the Deutsche Forschungsgemeinschaft (DFG, German Research Foundation) through projects A02, B02, B03, B07, and C03 of Collaborative Research Center SFB1242 ``Non-equilibrium dynamics of condensed matter in the time domain'' (Project ID 278162697).


\bibliography{Bibliography.bib}


\end{document}